\documentclass[prx,twocolumn,aps,amsmath,amssymb]{revtex4-2}
\usepackage[T1]{fontenc}
\usepackage[utf8]{inputenc}
\usepackage{fancyhdr}
\usepackage{graphicx}
\usepackage{amsmath}
\usepackage{amssymb}
\usepackage{dcolumn}
\usepackage{float}
\usepackage{bm}
\usepackage{subcaption}
\usepackage{amsfonts,times}
\usepackage{parskip}
\usepackage[breaklinks=true,colorlinks,citecolor=blue,linkcolor=blue,urlcolor=blue]{hyperref}

\DeclareMathAlphabet{\bi}{OML}{cmm}{b}{it}
\def\be{\begin{equation}}
\def\ee{\end{equation}}
\def\bearr{\begin{eqnarray}}
\def\eearr{\end{eqnarray}}

\begin{document}

\title{Spin-valley physics in anomalous thermoelectric responses of the spin-orbit coupled $\alpha$-$T_3$ system with broken time-reversal symmetry}

\author{Lakpa Tamang}
\author{Tutul Biswas}
\affiliation{Department of Physics, University of North Bengal, Raja Rammohunpur, Darjeeling-734013, India}
\email{tbiswas@nbu.ac.in}

\date{\today}

\begin{abstract}
We extract spin-valley physics in the anomalous Hall and Nernst responses of the $\alpha$-$T_3$ system, considering the simultaneous presence of the Kane–Mele type intrinsic spin–orbit interaction (SOI) and a time-reversal symmetry breaking staggered magnetization. Within a full low-energy continuum model, we compute the spin- and valley-resolved anomalous Hall, and anomalous Nernst conductivities. We show that the interplay between the SOI, magnetization, and a model parameter $\alpha$ for the $\alpha$-$T_3$ lattice enables efficient tuning of spin- and valley-dependent Hall and Nernst signals. The spin-valley physics of the Hall and Nernst responses in the absence and presence of the magnetization are well explained. The peak-dip features of the Nernst responses are also understood from the corresponding Hall responses through the Mott relation. We find that the magnetization introduces highly tunable spin and valley polarizations, which are calculated from the spin- and valley-resolved Nernst conductivities. It is shown that both the spin and valley polarizations can attain nearly complete polarization over extended regions of the parameter space. 
\end{abstract}

\maketitle

\section{Introduction}

The generation and control of spin-polarized currents constitute a solid foundation of modern spintronics. Spintronics, which exploits the intrinsic spin of electrons in addition to their charge, has emerged as a promising avenue for the realization of energy-efficient and multifunctional devices~\cite{Spin_T1,Spin_T2,Spin_T3,Spin_T4, Spin_T5}. Over the years, considerable effort has been devoted into developing efficient mechanisms for producing spin-polarized currents. These include electrical injection through ferromagnets~\cite{Spin_F1,Spin_F2}, the spin Hall effect~\cite{SHE_1,SHE_2,SHE_3}, spin injection~\cite{OPT_INJ_1,OPT_INJ_2,OPT_INJ_3}, voltage controllable generator~\cite{VOLT_1}, and various spin pumping schemes~\cite{PUMP_1,PUMP_2,PUMP_3,PUMP_4,PUMP_5,APUMP_1,APUMP_2,APUMP_3}. 

A closely related research direction, spin caloritronics~\cite{SPIN_C1,SPIN_C2}, explores the generation and control of spin currents through thermal gradients rather than applied electric fields. A central advance in this field is the observation of the spin Seebeck effect~\cite{SPIN_C3,SPIN_C4,SPIN_C5,SPIN_C6,SPIN_C7}, which demonstrates that a temperature gradient can be used as an efficient tool to generate spin current. Analogously, valley caloritronics~\cite{VALLEY_CAL1} combines thermoelectric transport with the valley degree of freedom emerging from two inequivalent valleys in the low-energy band structure of Dirac materials such as graphene~\cite{VALLEY_3, VALLEY_4, VALLEY_5}, silicene~\cite{VSILICENE}, transition-metal dichalcogenides~\cite{VALLEY_6, VTMD}, magnetic systems~\cite{2D_MAG1,2D_MAG2} etc. Electronic states near these valleys act as distinct pseudospin flavors, enabling their selective manipulation via external fields or symmetry-breaking perturbations.

The emergence of nontrivial Berry curvature in such materials due to some broken symmetries has attracted significant attention, leading to extensive studies of unconventional transport phenomena, including the anomalous Hall and Nernst effects, their spin and valley analogues, as well as the thermal Hall responses~\cite{AHall1,AHall2,SANE1,SANE2,SANE3,ATHE1_TH,ATHE2_TH}. Experimental demonstrations of these effects~\cite{VANE1_EX,ANE1, ANE2,ATHE1_EX,ATHE2_EX,ATHE3_EX,ATHE4_EX} further underscore their technological relevance. In particular, Nernst responses have played a pivotal role in identifying vortex phases in type-II superconductors~\cite{ANE1,ANE2}, probing topological surface states~\cite{ATHE1_TH}, and enabling the generation of pure spin and valley currents~\cite{SANE1}. 

Motivated by these developments, we focus on Berry curvature-mediated spin and valley transport in the
$\alpha$-$T_{3}$ system~\cite{THESIS}, a 
two-dimensional (2D) lattice model that serves as a link between graphene and dice lattice~\cite{dice1,dice2,dice3} through the smooth tuning of the parameter $\alpha$ between $0$ and $1$. Within nearest-neighbor tight-binding approximation, the low-energy spectrum of such a system possesses two inequivalent Dirac valleys $K$ and $K^\prime$. Each valley consists of a pair of linearly dispersive spin-degenerate bands accompanied by a dispersionless flat band at zero energy. Interestingly, the Berry phase corresponding to the $\alpha$-$T_3$ system depends on $\alpha$, which gives rise to a broad range of unusual phenomena including orbital magnetic susceptibility~\cite{Mag_Suscp}, quantized Hall response~\cite{T3_Hall1,T3_Hall2}, Klein tunneling~\cite{Klein2}, Weiss oscillations \cite{Weiss}, zitterbewegung~\cite{ZB}, unconventional plasmonic and optical features~\cite{Plasmon2,Plasmon3,Plasmon4,Mag_Opt1, Mag_Opt2, Mag_Opt3, Mag_Opt4}, Ruderman-Kittel-Kasuya-Yosida interaction~\cite{RKKY1,RKKY2}, minimal conductivity~\cite{ Min_Con}, thermoelectric effects~\cite{Nanoribbon_alpha, Thermo_Alpha1,Thermo_Alpha2}, modified Haldane model~\cite{alpha_Haldane}, etc.
Experimental routes to realizing $\alpha$-$T_3$ physics have been proposed in semiconductor heterostructures and cold-atom optical lattices~\cite{Dice_Real,Dice_Opt}. In particular, Hg$_{1-x}$Cd$_x$Te heterostructure at critical doping ($x=0.17$) has been demonstrated theoretically to mimic an $\alpha$-$T_3$ system with 
$\alpha=1/\sqrt{3}$~\cite{alp_T3_real}. In a series of recent works~\cite{{Flo1},{Flo2},{Flo3},{Flo4},{Flo5}}, it has been explicitly shown that an off-resonant circular polarized radiation can induce topological phases into the $\alpha$-$T_3$ model. Notably, the irradiated system exhibits a topological phase transition (TPT) across 
$\alpha=1/\sqrt{2}$, characterized by a quantized jump in the Chern number from $C=1$ to $C=2$.

The incorporation of next-nearest-neighbor hoppings in the tight-binding Hamiltonian gives rise to intrinsic spin-orbit interaction (SOI) of Kane–Mele type~\cite{Kane1,Kane2}, enabling the emergence quantum spin-Hall (QSH) phases in the $\alpha$-$T_3$ model. Moreover, the system undergoes a TPT across 
$\alpha=1/2$, accompanied by a change in the spin Chern number from $C_s=1$ to $C_s=2$~\cite{Spin_Hall_Phase}. In a recent study~\cite{Orbit_M}, the signatures of such a TPT have been identified in the orbital magnetization and circular dichroism. However, the symmetries of the underlying system impose severe constraints on the experimental observation of such effects. Introducing a staggered sublattice magnetization $M$, which breaks the time-reversal symmetry, lifts these restrictions~\cite{Spin_Hall_Phase, Orbit_M}. The interplay between $M$ and $\alpha$ also generates a rich phase diagram characterized by distinct combinations of $C$ and $C_s$~\cite{Spin_Hall_Phase}. A recent study also predicts a similar phase diagram when the spin-orbit coupled $\alpha$-$T_3$ system is exposed to off-resonant circularly polarized light~\cite{spin_T3_Flo}.

The thermoelectric properties of $\alpha$-$T_3$ systems, both in the presence and absence of a magnetic field have been investigated in the recent past~\cite{Nanoribbon_alpha,Thermo_Alpha1,Thermo_Alpha2}; however, the role of SOI has remained largely overlooked. To address this gap, here we investigate the anomalous thermoelectric response of the $\alpha$-$T_3$ system in the presence of intrinsic Kane-Mele type SOI. More specifically, we perform a detailed analysis of spin and valley Hall and Nernst effects in the spin-orbit coupled $\alpha$-$T_{3}$ system, both in the presence and absence of a staggered sublattice magnetization. 
Using a low-energy Kane–Mele–type model, we compute the Berry curvature and entropy-weighted transport coefficients. We show that the interplay of various parameters such as chemical potential, SOI strength, magnetization, and $\alpha$ enables precise control over spin- and valley-polarized thermoelectric responses. It is also demonstrated that the Nernst response of the $\alpha$-$T_3$ system produces tunable spin and valley polarizations, exhibits rich structures, featuring extended regions of nearly complete polarization.

The rest of the paper is organized as follows. 
In Sec.~\ref{Model}, we briefly introduce the essential features of the low-energy bandstructure of the $\alpha$-$T_{3}$ system, incorporating the intrinsic Kane–Mele SOI term and a staggered magnetization.
Section~\ref{Formalism} presents the formalism for Berry curvature driven anomalous thermoelectric transport focusing on the spin- and valley-resolved Hall and Nernst conductivities. 
In Sec.~\ref{Results}, we provide a detailed numerical analysis of the band dispersions, Berry curvature distributions, and the associated Hall and Nernst responses, as well as the spin and valley polarization, both in the presence and absence of the magnetization. Finally, Sec.~\ref{conclusion} summarizes our main findings.

\begin{figure}[h!]
\centering
\includegraphics[width=0.8\linewidth]{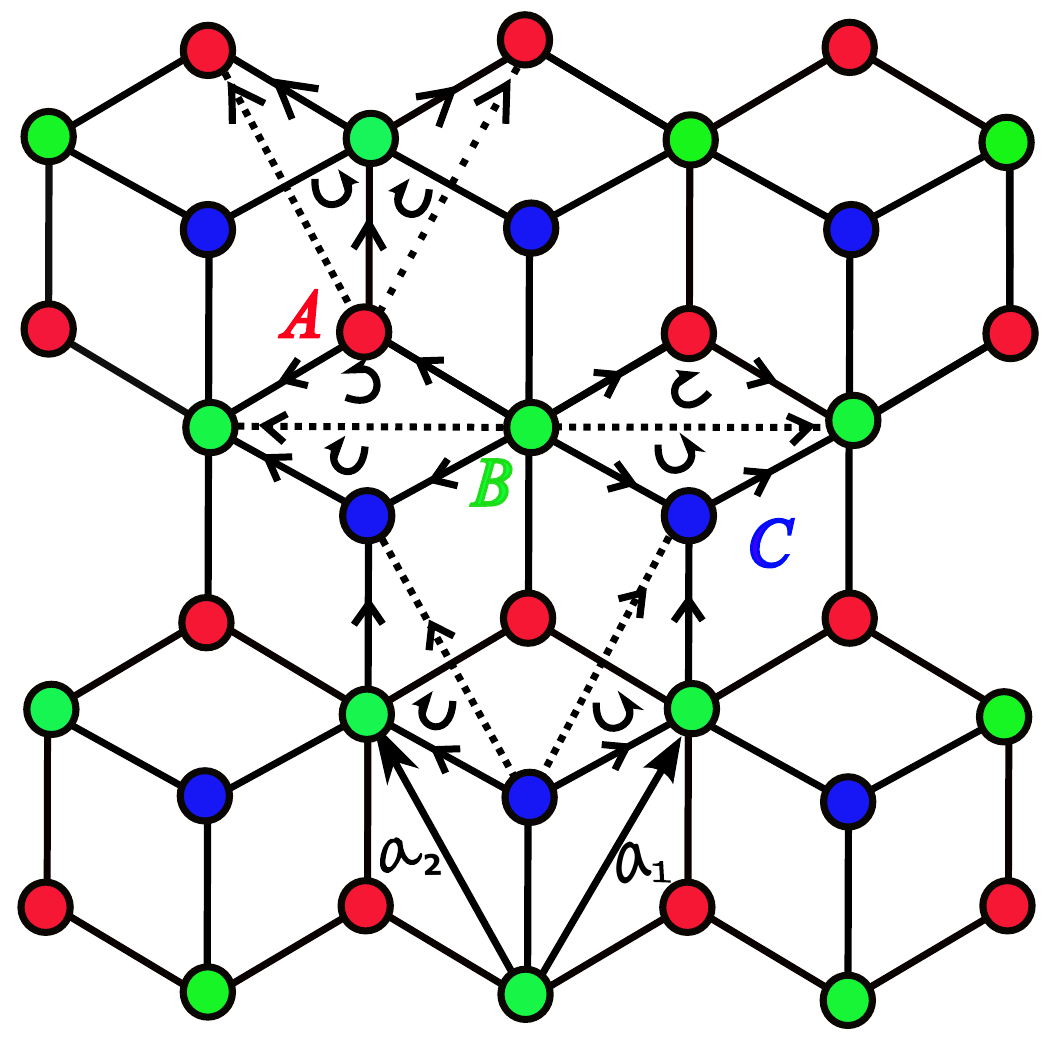}
\caption{(Color online) Schematic illustration of the $\alpha$–$T_{3}$ lattice with hub (B) and rim (A, C) sites. The NN (NNN) hopping paths are indicated by solid (dashed) lines. The NN B-C hopping amplitude is $\alpha$ times to that of A-B hopping strength. The lattice translational vectors are ${\bm a_1}=(\frac{\sqrt{3}}{2},\frac{3}{2})\rm a$ and ${\bm a_2}=(-\frac{\sqrt{3}}{2},\frac{3}{2})\rm a$, where $\rm a$ is the NN distance.}
\label{fig:lattice_structure}
\end{figure}

\section{Low-Energy description of the spin-orbit coupled $\alpha$-$T_3$ system} \label{Model}

We begin by briefly describing the spin-orbit coupled $\alpha$--$T_3$ lattice, considering the next-nearest-neighbor (NNN) interactions. The inclusion of NNN hopping introduces Kane-Mele type SOI~\cite{Kane1}. As schematically illustrated in Fig.~\ref{fig:lattice_structure}, a NNN hopping path comprises two consecutive nearest-neighbor (NN) segments, which may be traversed either clockwise or counterclockwise when viewed from the above along the direction perpendicular to the lattice plane. For instance, an electron can hop from a given $B$ site to one of its NNN $B$ sites either through an intermediate $A$  or $C$ site.
There are four distinct NNN hopping channels: (i) $A$–$B$–$A$, (ii) $B$–$A$–$B$, (iii) $B$–$C$–$B$, and (iv) $C$–$B$–$C$. The first two pathways [(i) and (ii)] give rise to SOI of strength $\lambda$, while the latter two [(iii) and (iv)] contribute an SOI of strength $\alpha\lambda$. This scaling is consistent with the basic assumption of the 
$\alpha$-$T_3$ model, wherein the NN hopping amplitude between $B$ and $C$ sites is $\alpha$ times of that between $A$ and $B$ sites. The $A$–$C$–$A$ NNN process is neglected, following the same logic that the $\alpha$-$T_3$ model excludes any direct $A$–$C$ NN coupling.

By incorporating both NN and NNN hopping terms, the low-energy tight-binding Hamiltonian in momentum space for a given spin index $\sigma$ and valley index $\eta$ can be expressed as~\cite{Spin_Hall_Phase, Orbit_M}
\begin{eqnarray}\label{Ham_eta}
H_\sigma^\eta(\bm{k}) = \lambda \eta \sigma
\begin{pmatrix}
-\cos\phi & f_{\bm{k}} \cos\phi & 0 \\
f_{\bm{k}}^* \cos\phi & \cos\phi - \sin\phi & f_{\bm{k}} \sin\phi \\
0 & f_{\bm{k}}^* \sin\phi & \sin\phi
\end{pmatrix},
\end{eqnarray}
where $\phi = \arctan(\alpha)$ and 
$f_{\bm{k}} = \sigma \hbar v_F (k_x + i \eta k_y)/\lambda$,
with $v_F$ denoting the Fermi velocity. The diagonal elements in the Hamiltonian indicate that the SOI induces an $\alpha$-dependent mass term in the energy spectrum.

To incorporate the effect of staggered magnetization, we consider a simple $A$–$C$ sublattice configuration, where magnetization takes opposite signs on the $A$ and $C$ sites and vanishes on the $B$ sites. It is important to note that this $A$–$C$ staggered configuration is specific to the $\alpha$–$T_3$ lattice with $\alpha \neq 0$. In the graphene limit ($\alpha = 0$), the $C$ sublattice is absent, and the lattice reduces to the standard honeycomb structure consisting only of $A$ and $B$ sublattices. Consequently, for $\alpha = 0$ one must instead consider staggered magnetization between the $A$ and $B$ sublattices. Thus, we adopt an $A$–$C$ staggered configuration for $\alpha \neq 0$, and an $A$–$B$ staggered configuration in the graphene limit.

Therefore, for $\alpha\neq0$, the staggered magnetization term added to the Hamiltonian in Eq.~(\ref{Ham_eta}) reads as~\cite{Spin_Hall_Phase, Orbit_M}
\begin{equation}
H_{M} = M \sigma s^z = M \sigma
\begin{pmatrix}
1 & 0 & 0 \\
0 & 0 & 0 \\
0 & 0 & -1
\end{pmatrix},
\label{HM}
\end{equation}
where $s^z$ represents the $z$-component of the pseudospin-$1$ operator associated with the lattice. The presence of $H_{M}$ explicitly breaks the time-reversal symmetry (TRS) of the Hamiltonian.

The energy spectrum is then obtained as \cite{Orbit_M} (see Appendix. \ref{Appen_A} for derivation)
\begin{equation}\label{Energy_SOI}
\varepsilon_{\eta,\sigma}^n(\bm{k}) = 2 \sqrt{\frac{-p}{3}} 
\cos\!\left[\frac{1}{3}\arccos\!\left(\frac{3q}{2p}\sqrt{\frac{-3}{p}}\right) 
- \frac{2\pi n}{3}\right],
\end{equation}
with
\begin{equation}
p = \frac{1}{2}\lambda^2 \sin(2\phi)+\sqrt{2}\lambda M\eta\sin\!\left(\phi + \frac{\pi}{4}\right) - (\lambda^2 + \hbar^2 v_F^2 k^2+M^2),
\end{equation}
and
\begin{equation}
\begin{aligned}
q ={}&
\frac{1}{\sqrt{2}} \lambda \sigma \eta \sin(2\phi)
\cos\!\left(\phi + \frac{\pi}{4}\right)
\left(\lambda^2 + \hbar^2 v_F^2 k^2\right) \\
&+ \sqrt{2}\,\lambda \sigma \eta
\cos\!\left(\phi + \frac{\pi}{4}\right) {M^2} \\
&- M \sigma \cos(2\phi)
\left(\lambda^2 + \hbar^2 v_F^2 k^2\right).
\end{aligned}
\end{equation}

The corresponding eigenfunction is expressed as
\begin{equation}\label{Eig_SOI}
\psi_{\eta,\sigma}^n(\bm{k}) = 
\frac{1}{\sqrt{1 + |F_{\bm{k}}|^2 + |G_{\bm{k}}|^2}}
\begin{pmatrix}
F_{\bm{k}} \\ 1 \\ G_{\bm{k}}
\end{pmatrix}
\frac{e^{i\bm{k}\cdot\bm{r}}}{\sqrt{S}},
\end{equation}
where
$F_{\bm{k}} =f_{\bm{k}} \cos\phi/(\tilde{\varepsilon}_{\eta,\sigma}^n + \cos\phi)$ and 
$G_{\bm{k}} = f_{\bm{k}}^* \sin\phi/(\tilde{\varepsilon}_{\eta,\sigma}^n - \sin\phi)$
with $\tilde{\varepsilon}_{\eta,\sigma}^n = \varepsilon_{\eta,\sigma}^n / (\eta \sigma \lambda)$. Here, $S$ is the area of the sample.

The quantity $n$ in Eqs. \eqref{Energy_SOI} and \eqref{Eig_SOI} denotes the band index, with
 $n = 0, 1, 2$ corresponding to the conduction band (CB), flat band (FB), and valence band (VB), respectively. In the absence of $M$, the TRS is preserved for all $\alpha$, whereas the inversion symmetry (IS) is broken for $0 < \alpha < 1$. Therefore, in the limiting cases 
 $\alpha = 0$ (graphene) and $\alpha = 1$ (dice lattice), both the TRS and IS are simultaneously present, resulting in spin and valley degenerate energy bands.  
However, the presence of the staggered magnetization lifts only the valley degeneracy of the energy bands in these two limiting cases. Nevertheless, for intermediate values of $\alpha$, both the spin and valley degeneracies are lifted due to the simultaneous breaking of IS and TRS. 
For $\alpha=1$, the expression for the energy eigenvalue simplifies to 
\begin{equation}
  \varepsilon^{\rm d}=\left(\hbar^2 v_F^2k^2 + \Delta_{\rm d}^2\right)^{1/2}  
\end{equation}
where, $\Delta_{\rm d}=(\lambda\eta\sigma/\sqrt{2}-M\sigma)$ and similarly, for $\alpha=0$ the energy eigenvalue can be expressed as 
\begin{equation}
\varepsilon^{\rm g}=\left(\hbar^2 v_F^2k^2 + \Delta_{\rm g}^2\right)^{1/2}    
\end{equation}
with $\Delta_{\rm g}=(\lambda\eta\sigma-M\sigma)$.

\section{Formalism for the Berry curvature mediated anomalous thermoelectric transport }\label{Formalism}
The anomalous Hall (Nernst) effect refers to the generation of a transverse charge current in response to a longitudinal electric field (temperature gradient) in systems possessing finite Berry curvature. The Berry curvature acts as an effective magnetic field in momentum space and modifies the carrier dynamics by introducing an anomalous contribution to the electron velocity. In the following, we briefly outline the essential theoretical framework underlying anomalous thermoelectric transport.

In the linear-response regime, the charge current density ${\bm J}$ and the heat current density 
${\bm J}^h$ are related to the externally applied electric field ($\bm E$) and temperature gradient ($-{\bm \nabla} T$) through the equation 
\begin{eqnarray}\label{Ther_eq}
\begin{pmatrix}
    \bm J\\
    \bm J^h
\end{pmatrix}
=\begin{pmatrix}
    \widetilde{\bm\sigma} & \widetilde{\bm \alpha}\\
    \widetilde{\bar{\bm\alpha}} & \widetilde{\bm l}
\end{pmatrix}
\begin{pmatrix}
    \bm E\\
    -\bm\nabla T
\end{pmatrix}.
\end{eqnarray}
Here, $\widetilde{\bm \sigma}$ and $\widetilde{\bm\alpha}$ are the charge conductivity tensor and thermoelectric tensor, respectively. The tensor $\widetilde{\bm l}$ connects the heat current to the temperature gradient. In the absence of an electric field ($\bm E=0$), $\widetilde{\bm l}$ becomes the thermal conductivity tensor. Onsager's reciprocity relation further dictates that 
$\widetilde{\bar{\bm \alpha}}=T\widetilde{\bm\alpha}$. Eq.~\eqref{Ther_eq} may alternatively be rewritten componentwise as
\begin{eqnarray}
    J_{\mu\nu}=\sigma_{\mu\nu}E_\nu+\alpha_{\mu\nu}(-\partial_\nu T)\nonumber\\
    J_{\mu\nu}^h=\bar{\alpha}_{\mu\nu}E_\nu+l_{\mu\nu}(-\partial_\nu T),
\end{eqnarray}
where $\mu,\nu=x,y$ label the spatial coordinates and summation over repeated indices is implied.

In the transverse transport geometry, an electric field applied along 
the $y$-direction generates a charge current along the $x$-direction according to $J_x=\sigma_{xy}E_y$, where $\sigma_{xy}$ denotes the anomalous Hall conductivity (AHC). Analogously, a temperature gradient along the $y$-direction induces a transverse charge current along the 
$x$-direction, described by $J_x=\alpha_{xy}(-\partial_yT)$, where $\alpha_{xy}$ represents the anomalous Nernst conductivity (ANC).

To obtain the expressions for the AHC and ANC, we employ the
semiclassical dynamics of Bloch electrons in the presence of Berry curvature. 
For an electron in a given Bloch band $(n, \eta, \sigma)$, the modified velocity in the presence of an external electric field reads \cite{Berry_Niu, Berry_Sund}
\begin{equation}
\bm{v}_{\eta,\sigma}^n(\bm{k}) =
\frac{1}{\hbar} \bm{\nabla}_{\bm{k}} \varepsilon_{\eta,\sigma}^n(\bm{k})
- \frac{e}{\hbar} \bm{E} \times \bm{\Omega}_{\eta,\sigma}^n(\bm{k}),
\end{equation}
where 
\begin{equation}\label{Berry}
\bm{\Omega}_{\eta,\sigma}^n(\bm{k}) =
\bm{\nabla}_{\bm{k}} \times 
\langle u_{\eta,\sigma}^n(\bm{k}) | i \bm{\nabla}_{\bm{k}} | u_{\eta,\sigma}^n(\bm{k}) \rangle
\end{equation}
is the Berry curvature associated with the periodic part of the Bloch function
$|u_{\eta,\sigma}^n(\bm{k})\rangle = |\psi_{\eta,\sigma}^n(\bm{k})\rangle e^{-i\bm{k}\cdot\bm{r}}\sqrt{S}$.
The symmetry properties of 
${\Omega}_{\eta,\sigma}^n(\bm k)$ imply that a nonzero Berry curvature requires the breaking of either TRS or IS or both.
In a two-dimensional (2D) system confined to the $xy$ plane, only the out-of-plane ($z$) component ${\Omega}_{\eta,\sigma}^n(\bm{k})$ of the Berry curvature contributes.   
For an electric field applied along the 
$y$-direction, the transverse anomalous velocity in the $x$-direction becomes
$v_x^{\rm an}(\bm k)=-eE_y\Omega_{\eta,\sigma}^n(\bm k)/\hbar$.

The spin- and valley-resolved anomalous Hall current is obtained by summing over occupied states,
\begin{eqnarray}\label{hall_current}
J_x^{\eta,\sigma}
=
-e\sum_n\int \frac{d^2k}{(2\pi)^2}
\,v_x^{\rm an}(\bm k)\,f_{\eta,\sigma}^n(\bm k),
\end{eqnarray}
where $f_{\eta,\sigma}^n(\bm{k}) = [e^{\beta(\varepsilon_{\eta,\sigma}^n - \mu)} + 1]^{-1}$ is usual Fermi–Dirac distribution with $\beta = 1/(k_B T)$ and $\mu$ as the chemical potential.
Substituting $v_x^{\rm an}(\bm k)$ into
Eq.~(\ref{hall_current}) yields
\begin{equation}
J_x^{\eta,\sigma}
=
\frac{e^2}{\hbar}E_y
\sum_n\int
\frac{d^2k}{(2\pi)^2}
\Omega_{\eta,\sigma}^n(\bm k)
f_{\eta,\sigma}^n(\bm k).
\end{equation}
Comparing with the definition
$J_x^{\eta,\sigma}=\sigma_{xy}E_y=\sigma_{\rm H}^{\eta,\sigma}E_y$,
the spin- and valley-resolved AHC is obtained as
\begin{eqnarray}\label{ahc}
\sigma_{\rm H}^{\eta,\sigma}
=
\frac{e^2}{\hbar}\sum_n
\int
\frac{d^2k}{(2\pi)^2}
\Omega_{\eta,\sigma}^n(\bm k)
f_{\eta,\sigma}^n(\bm k).
\end{eqnarray}
Equation~(\ref{ahc}) shows that the AHC is determined by the Berry curvature weighted by the occupation probability of the electronic states. Consequently, the AHC is fundamentally a Fermi-sea property.

The ANC can be derived in an analogous manner by considering entropy transport.
The entropy density associated with a fermionic state is
given by
\begin{eqnarray}\label{eq:entropy}
S_{\eta,\sigma}^n(\bm k)
=
&-&f_{\eta,\sigma}^n(\bm k)\ln f_{\eta,\sigma}^n(\bm k)\nonumber\\
&-&
[1-f_{\eta,\sigma}^n(\bm k)]\ln[1-f_{\eta,\sigma}^n(\bm k)].
\end{eqnarray}

The spin- and valley-resolved entropy current density generated by the anomalous velocity is
therefore
\begin{eqnarray}\label{eq:entropy_current}
J_x^{\eta,\sigma,S}
=
k_B\sum_n
\int
\frac{d^2k}{(2\pi)^2}
v_x^{\rm an}\,S_{\eta,\sigma}^n(\bm k),
\end{eqnarray}
which gives
\begin{eqnarray}
J_x^{\eta,\sigma,S}
=
-\frac{ek_B}{\hbar}E_y\sum_n
\int
\frac{d^2k}{(2\pi)^2}
\Omega_{\eta,\sigma}^n(\bm k)
S_{\eta,\sigma}^n(\bm k).
\end{eqnarray}

Since the heat current density is related to the entropy current density through $J_x^{\eta,\sigma,h}=TJ_x^{\eta,\sigma, S}$,
one obtains
\begin{eqnarray}\label{eq:heat_current}
J_x^{\eta,\sigma, h}
=
-\frac{Tek_B}{\hbar}E_y\sum_n
\int
\frac{d^2k}{(2\pi)^2}
\Omega_{\eta,\sigma}^n(\bm k)
S_{\eta,\sigma}^n(\bm k).
\end{eqnarray}

Comparing with the definition
$J_x^{\eta,\sigma, h}=\bar{\alpha}_{xy}E_y$,
and using the Onsager relation
$\bar{\alpha}_{xy}=T\alpha_{xy}=T\alpha_{\rm N}^{\eta,\sigma}$, the spin- and valley-resolved ANC is finally obtained as \cite{ATHE2_TH, Berry_Rev}
\begin{eqnarray}\label{eq:anc}
\alpha_{\rm N}^{\eta,\sigma}
=
-\frac{ek_B}{\hbar}\sum_n
\int
\frac{d^2k}{(2\pi)^2}
\Omega_{\eta,\sigma}^n(\bm k)
S_{\eta,\sigma}^n(\bm k).
\end{eqnarray}

Unlike the AHC, which depends on all occupied states, the ANC is weighted by the entropy density and is therefore dominated by states near the Fermi energy where the occupation probability changes rapidly.
Consequently, the ANC is highly sensitive to the electronic structure around the chemical potential and to particle-hole asymmetry, making it essentially a Fermi-surface property. Equation~\eqref{eq:anc} can also be expressed as \cite{ATHE2_TH}
\begin{eqnarray}\label{anc_alt}
\alpha_{\rm N}^{\eta,\sigma}=-\frac{1}{eT}\int d\varepsilon\, \frac{\partial f(\varepsilon)}{\partial\mu}\sigma_{\rm H, 0}^{\eta,\sigma}(\varepsilon)\,(\varepsilon-\mu),   
\end{eqnarray}
where $\sigma_{\rm H,0}^{\eta,\sigma}(\varepsilon)$ is the AHC at zero temperature, defined as
\begin{eqnarray}
\sigma_{\rm H, 0}^{\eta,\sigma}(\varepsilon)=\frac{e^2}{\hbar}\sum_n\int\frac{d^2k}{(2\pi)^2}\Theta[\varepsilon-\varepsilon_{\eta,\sigma}^n(\bm k)]\,\Omega_{\eta,\sigma}^n(\bm k).
\end{eqnarray}
 
Because the TRS enforces cancellation of the total (spin and valley summed) AHC and ANC for $M=0$, we focus on their spin and valley components, defined as
\begin{eqnarray}\label{SVHC}
\sigma_{\rm H}^{\rm s} = \sum_{\eta,\sigma} \sigma \, \sigma_{\rm H}^{\eta,\sigma},
\qquad
\sigma_{\rm H}^{\rm v} = \sum_{\eta,\sigma} \eta \, \sigma_{\rm H}^{\eta,\sigma},
\end{eqnarray}
and
\begin{eqnarray}\label{SVNC}
\alpha_{\rm N}^{\rm s} = \sum_{\eta,\sigma} \sigma \, \alpha_{\rm N}^{\eta,\sigma},
\qquad
\alpha_{\rm N}^{\rm v} = \sum_{\eta,\sigma} \eta \, \alpha_{\rm N}^{\eta,\sigma}.
\end{eqnarray}
Here, $\sigma_{\rm H}^{\rm s}$ and $\sigma_{\rm H}^{\rm v}$ denote spin Hall Conductivity (SHC) and valley Hall conductivity (VHC), respectively. In a similar way, 
$\alpha_{\rm N}^{\rm s}$ and $\alpha_{\rm N}^{\rm v}$ represent the spin Nernst conductivity (SNC) and valley Nernst conductivity (VNC), respectively.

For completeness, we introduce the spin- and valley-resolved polarization factors associated with the anomalous Nernst conductivity (ANC). These quantities provide a compact measure of the degree to which the transverse thermoelectric response originates from a single spin or valley channel.  We define the spin (valley) polarization of the ANC as \cite{SANE1}
\begin{equation}
P_{\rm s(v)} =
\frac{\left( \alpha^{K\uparrow}_{\rm N} + \alpha^{K'\uparrow (K\downarrow)}_{\rm N} \right)
- \left(\alpha^{K\downarrow(K'\uparrow)}_{\rm N} + \alpha^{K'\downarrow}_{\rm N} \right)}
{\left( |\alpha^{K\uparrow}_{\rm N}| + |\alpha^{K'\uparrow}_{\rm N}| 
+ |\alpha^{K\downarrow}_{\rm N}| + |\alpha^{K'\downarrow}_{\rm N}| \right)} ,
\label{eq:polarization}
\end{equation}
where $\alpha^{\eta,\sigma}_{\rm N}$ denotes the ANC contribution from valley $\eta=\{K,K'\}$ and spin $\sigma=\{\uparrow,\downarrow\}$.

A value $P_{\rm s(v)}=\pm 1$ indicates that the ANC originates entirely from a single spin (valley) channel, corresponding to a perfectly spin- or valley-polarized Nernst signal. Intermediate values $|P_{\rm s(v)}| < 1$ quantify partial polarization arising from mixed spin or valley contributions, and $P_{\rm s(v)}=0$ signifies that the Nernst response arises from nearly symmetric contributions of opposite spins (valleys).

\section{Results and discussions}\label{Results}
\subsection{Band structure and Berry curvature}
\begin{figure}[h!]
\centering
\includegraphics[width=9.7 cm, height=7.2cm]{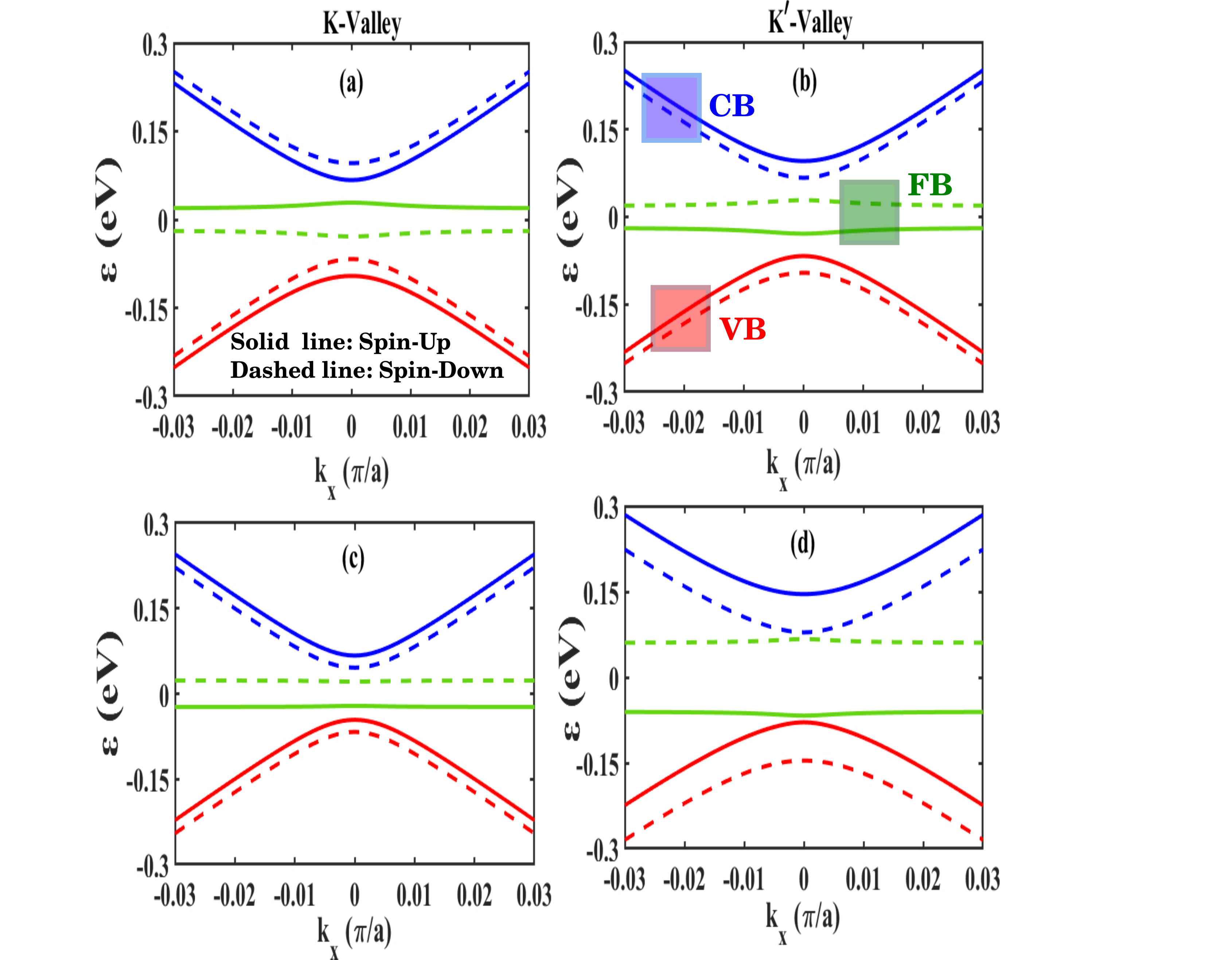}
\caption{(Color online) Low-energy dispersions in the $K$ and $K^\prime$ valleys for 
$\alpha=0.3$ and $\lambda=100$\,meV. Upper panels [(a) and (b)] represent $M=0$, while lower panels [(c) and (d)] correspond to $M\neq 0$ case.}
\label{fig:Energy}
\end{figure}

The low-energy band dispersions of the spin-orbit coupled $\alpha$-$T_3$ lattice, both in the absence and presence of a staggered magnetization, are shown in Fig.~\ref{fig:Energy}. Figures~\ref{fig:Energy}(a) and \ref{fig:Energy}(b) refer to the energy dispersions around the $K$ and $K^{\prime}$ valleys, respectively, for $M=0$, while Figs.~\ref{fig:Energy}(c) and \ref{fig:Energy}(d) represent the same for $M=50$~meV. The intrinsic SOI lifts the spin-degeneracy of the CB, FB, and VB by splitting them into their spin-up and spin-down counterparts. It also causes a distortion in the FB near the valley extremum. When $M=0$, the role of the spin-up sub-bands in the $K$ valley is interchanged by the spin-down sub-bands in the $K^\prime$ valley and vice versa, i.e., 
$\varepsilon_{K,\uparrow}=\varepsilon_{K^{\prime},\downarrow}$ and 
$\varepsilon_{K,\downarrow}=\varepsilon_{K^{\prime},\uparrow}$, owing to the TRS. 

When a finite staggered magnetization is introduced, the exact correspondence between the $K$ and $K^{\prime}$ valleys is lost as a result of broken TRS, leading to a significant modification of the band structure. The interplay of the magnetization and SOI leads to an enhanced spin-splitting in the $K^\prime$ valley [Fig.~\ref{fig:Energy}(d)] relative to the $K$ valley [Fig.~\ref{fig:Energy}(c)]. This asymmetry in the spin-splitting between the valleys has a pronounced effect on the thermoelectric response, as we will see later.

\begin{figure}[h!]
\centering
\includegraphics[width=8.5 cm, height=7.5cm]{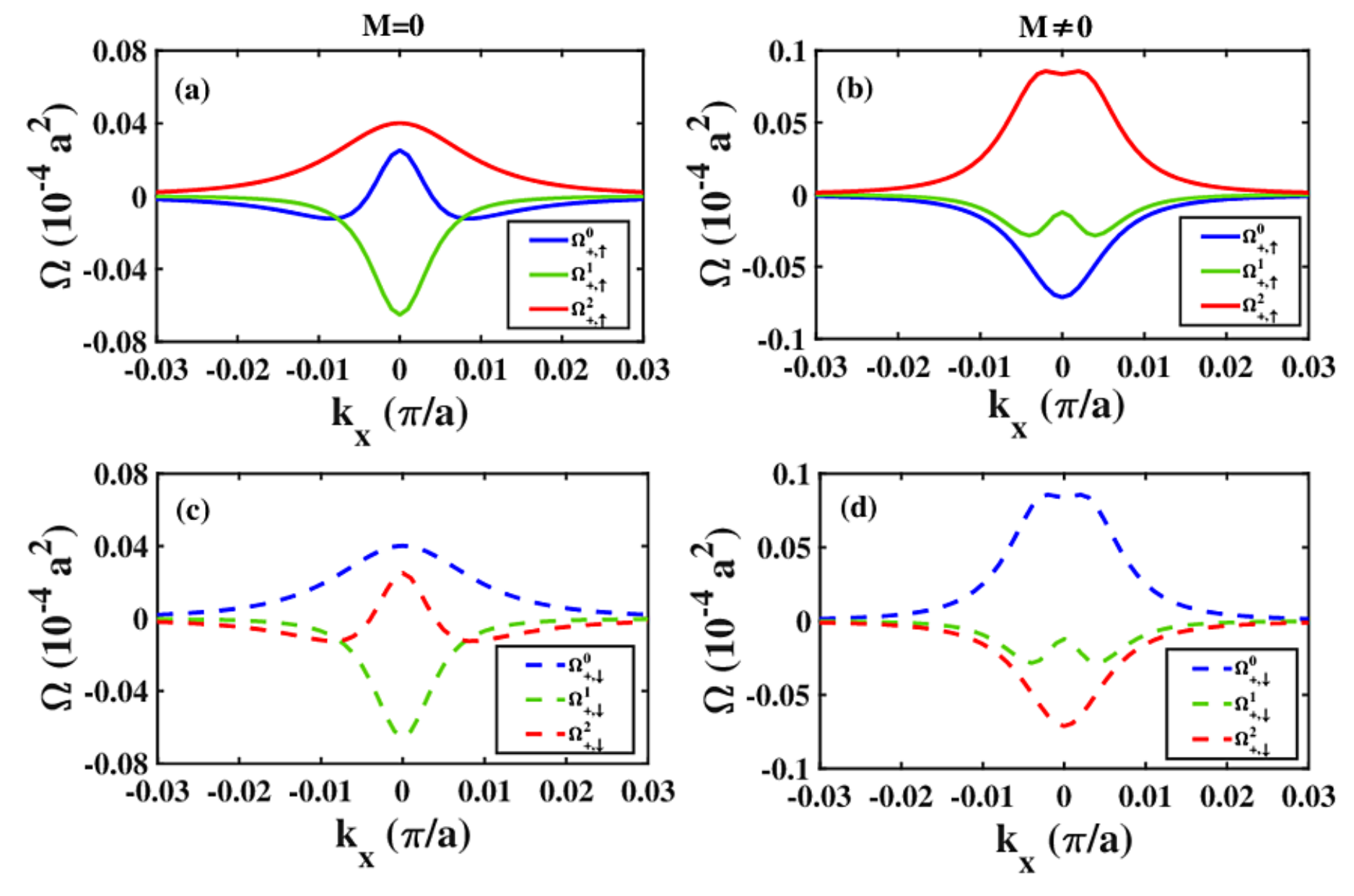}
\caption{(Color online) Berry curvature distribution $\Omega^{n}_{K,\sigma}(\mathbf{k})$ in the $K$ valley for spin-up and spin-down bands. Left Panels [(a) and (c)] are for $M=0$, while right panels [(b) and (d)] represents $M\neq 0$ case.}
\label{fig:BerryK}
\end{figure}

\begin{figure}[h!]
\centering
\includegraphics[width=8.5 cm, height=7.5 cm]{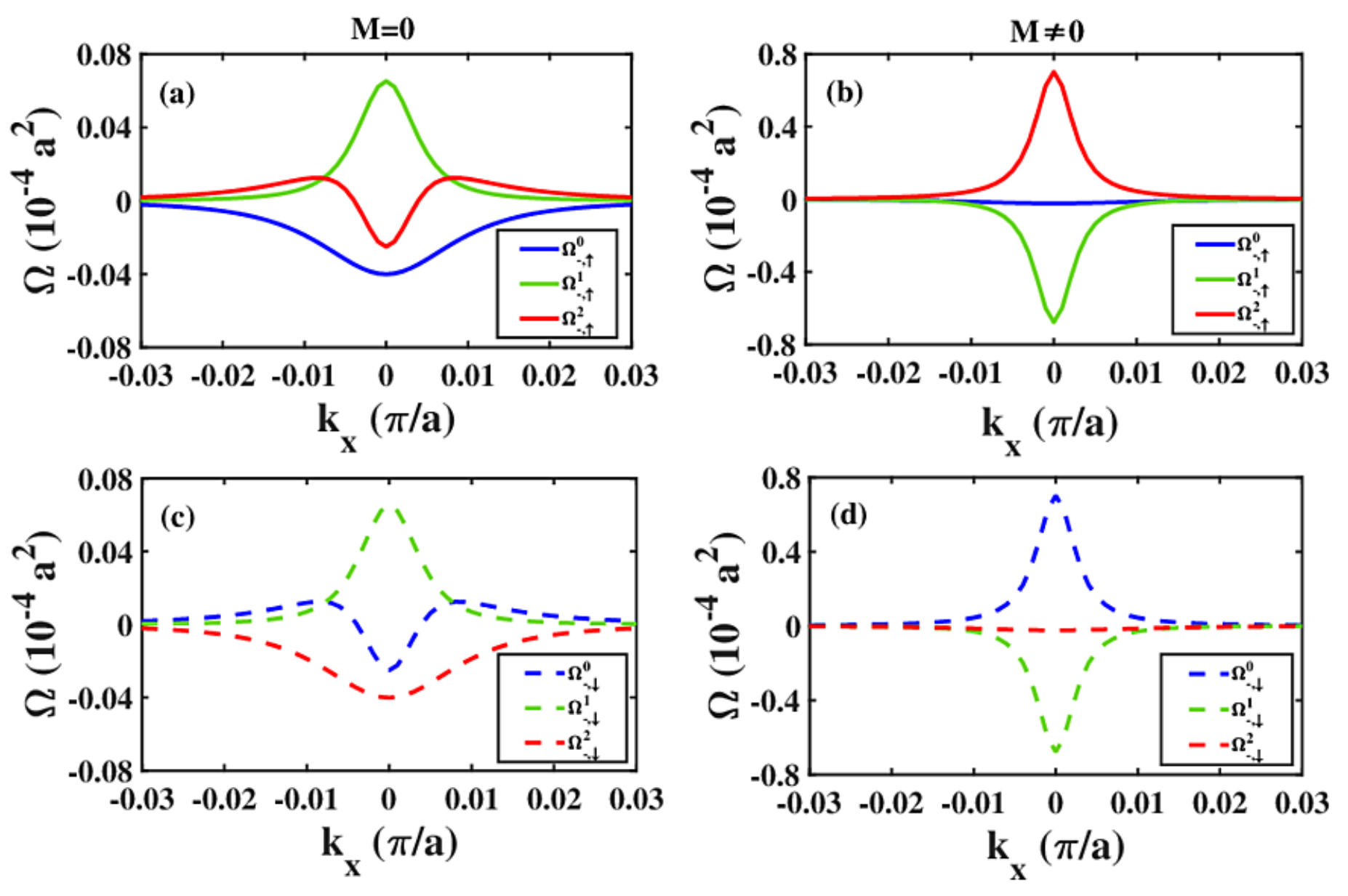}
\caption{(Color online) Same as Fig. \ref{fig:BerryK} but for the $K^{\prime}$ valley.}
\label{fig:BerryKP}
\end{figure}

Using Eq. (\ref{Berry}), we numerically calculate the Berry curvature for intermediate values of $\alpha$ both in the absence and presence of the staggered magnetization. The distributions of the Berry curvature for both spin orientations ($\sigma = \uparrow, \downarrow$) in the $K$ and $K^{\prime}$ valleys are depicted in Figs.~\ref{fig:BerryK} and~\ref{fig:BerryKP}, respectively. For both cases $M=0$ and $M\neq 0$, the Berry curvatures of all spin-split sub-bands are predominantly concentrated near the Dirac points and exhibit clear spin–valley resolution. The FB acquires a finite yet spin-degenerate Berry curvature, satisfying $\Omega^{1}_{\eta,\sigma}(\bm{k}) = \Omega^{1}_{\eta,-\sigma}(\bm{k})$. However, the Berry curvatures associated with the CB and VB are spin-polarized, satisfying the relation:
$\Omega^{0}_{\eta,\sigma}(\bm{k}) = \Omega^{2}_{\eta,-\sigma}(\bm{k})$.

In the absence of magnetization ($M=0$), the Berry curvature obeys
$\Omega^{n}_{\eta,\sigma}(\bm{k}) = -\,\Omega^{n}_{-\eta,-\sigma}(\bm{k})$, guaranteed by the TRS. This relation indicates that the total Berry curvature summed over the spin and valley degrees of freedom 
is zero, resulting in a vanishing net anomalous Hall response, despite significant local Berry curvature concentrated around the valleys.

The broken TRS for $M\neq0$ leads to a significant redistribution and enhancement of the Berry curvature across all spin-split subbands.
The staggered magnetization modifies the effective band gap, producing opposite effects in the $K$ and $K^\prime$ valleys. Consequently, the Berry curvature is enhanced in the valley with the reduced band gap and suppressed in the valley with the enlarged band gap. An analytical derivation of the Berry curvature at 
$\bm{k}=0$ for a given $\alpha$ is presented in Appendix B, which explicitly demonstrates how the interplay between SOI and staggered magnetization alters the band gaps in the two valleys.
Specifically, the Berry curvatures associated with the CB and VB gain a dominant spin-dependent curvature near the valleys. The Berry curvature in the $K^\prime$ valley no longer mirrors that of the $K$ valley, resulting in a finite imbalance between the spin channels. As a result, the spin and valley summed Berry curvature becomes nonzero, leading to a net anomalous Hall response. The pronounced contrast in Berry curvature patterns for different spin channels across both valleys highlights the tunability of spin and valley degrees of freedom via variation of the staggered magnetization.

Additionally, we obtain the following analytical expressions of the Berry curvature 
\begin{equation}\label{Berry_dice}
    \Omega^{n}_{\eta,\sigma} (\bm k)=
\frac{\eta(n-1)\Delta_{\rm d} \hbar^2 v_F^2}
{\left[\hbar^2 v_F^2k^2 + \Delta_{\rm d}^2\right]^{3/2}} 
\end{equation}
for $\alpha=1$
and 
\begin{equation}\label{Berry_graph}
    \Omega^{n}_{\eta,\sigma} (\bm k)=
\frac{\eta(n-1)\Delta_{\rm g} \hbar^2 v_F^2}
{2\left[\hbar^2 v_F^2k^2 + \Delta_{\rm g}^2\right]^{3/2}} 
\end{equation}
for $\alpha=0$.
The index $n$ takes only $0$ and $2$ in Eq. \eqref{Berry_graph}, representing the CB and VB of graphene, respectively. It is evident from Eqs. (\ref{Berry_dice}) and (\ref{Berry_graph}) that $\Omega_{\eta,\sigma}$ is both spin and valley resolved, as $\Delta_{\rm d}$ and $\Delta_{\rm g}$ are spin and valley dependent.  

\subsection{Valley and spin Hall conductivities}
In this section, we systematically present a detailed analysis of the VHC and SHC, computed numerically using Eqs.~(\ref{ahc}) and~(\ref{SVHC}). The temperature is fixed at $T=20$ K.

In the following, we discuss two physically distinct regimes: the TRS-preserving case ($M=0$) and the TRS-broken case ($M\neq 0$).

\begin{figure}[H]
\centering
\includegraphics[width=8.5 cm, height=4.5 cm]{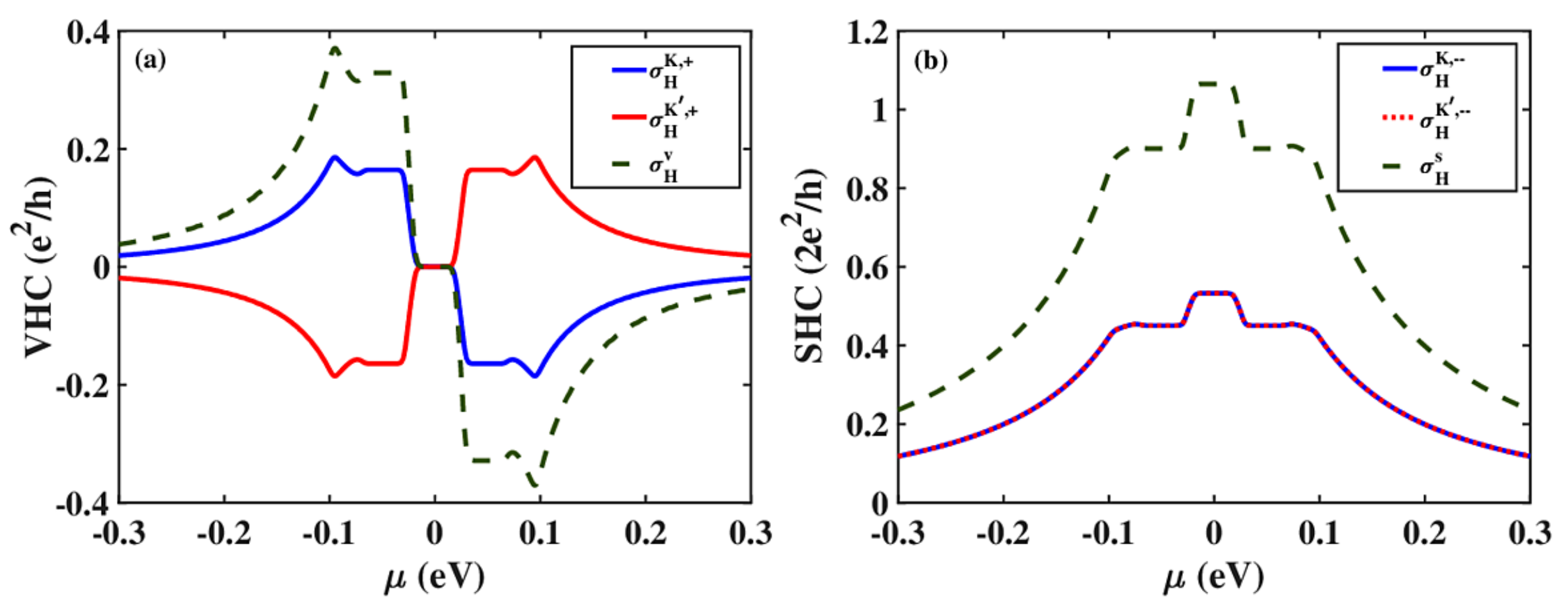}
\caption{(Color online) (a) Valley Hall conductivity and (b) Spin Hall conductivity as a function of chemical potential 
$\mu$ for  $M=0$, $\alpha=0.3$, $T=20$ K, and 
$\lambda=100$ meV.}
\label{fig:svh_salp}
\end{figure}

\subsubsection {Case I: Hall conductivities in the absence of the staggered magnetization ($M = 0$)}

In this case, the anomalous Hall response is solely determined by the Berry curvature generated by the SOI. In Fig.~\ref{fig:svh_salp}(a), we show the variation of the Hall conductivity for a given valley summed over the spin index, i.e., 
$\sigma_{\rm H}^{\eta,+}=\sigma_{\rm H}^{\eta,\uparrow}+\sigma_{\rm H}^{\eta,\downarrow}$, along with the VHC, as functions of the chemical potential $\mu$, considering $\alpha=0.3$.
The spin integrated AHC in the $K$ valley has the same magnitude but opposite sign compared to that in the $K^\prime$ valley due to the TRS. Therefore, the total AHC vanishes. However, a pure VHC, i.e., $\sigma_{\rm H}^{\rm v}=\sigma_{\rm H}^{K,+}-\sigma_{\rm H}^{K^\prime,+}$ persists, as shown by the dashed line in Fig.~\ref{fig:svh_salp}(a). In addition, the VHC changes sign upon reversal of the chemical potential, demonstrating antisymmetry in $\mu$. Figure~\ref{fig:svh_salp}(b) shows the SHC as a function of $\mu$. For a given valley $\eta$, the SHC is defined as $\sigma_{\rm H}^{\eta,-}=\sigma_{\rm H}^{\eta,\uparrow}-\sigma_{\rm H}^{\eta,\downarrow}$, while the total SHC is given by $\sigma_{\rm H}^{\rm s}=\sigma_{\rm H}^{K,-}+\sigma_{\rm H}^{K^\prime,-}$. The SHC contributions from the $K$ and $K^\prime$ valleys are identical, leading to a nonvanishing total SHC. Unlike the VHC, the SHC remains invariant under the transformation $\mu \to -\mu$. Both the VHC and SHC display plateau structures in three distinct ranges of $\mu$, which can be identified as SOI induced energy gaps.

\begin{figure}[t] 
\centering
\includegraphics[width=8.5cm, height=6cm]{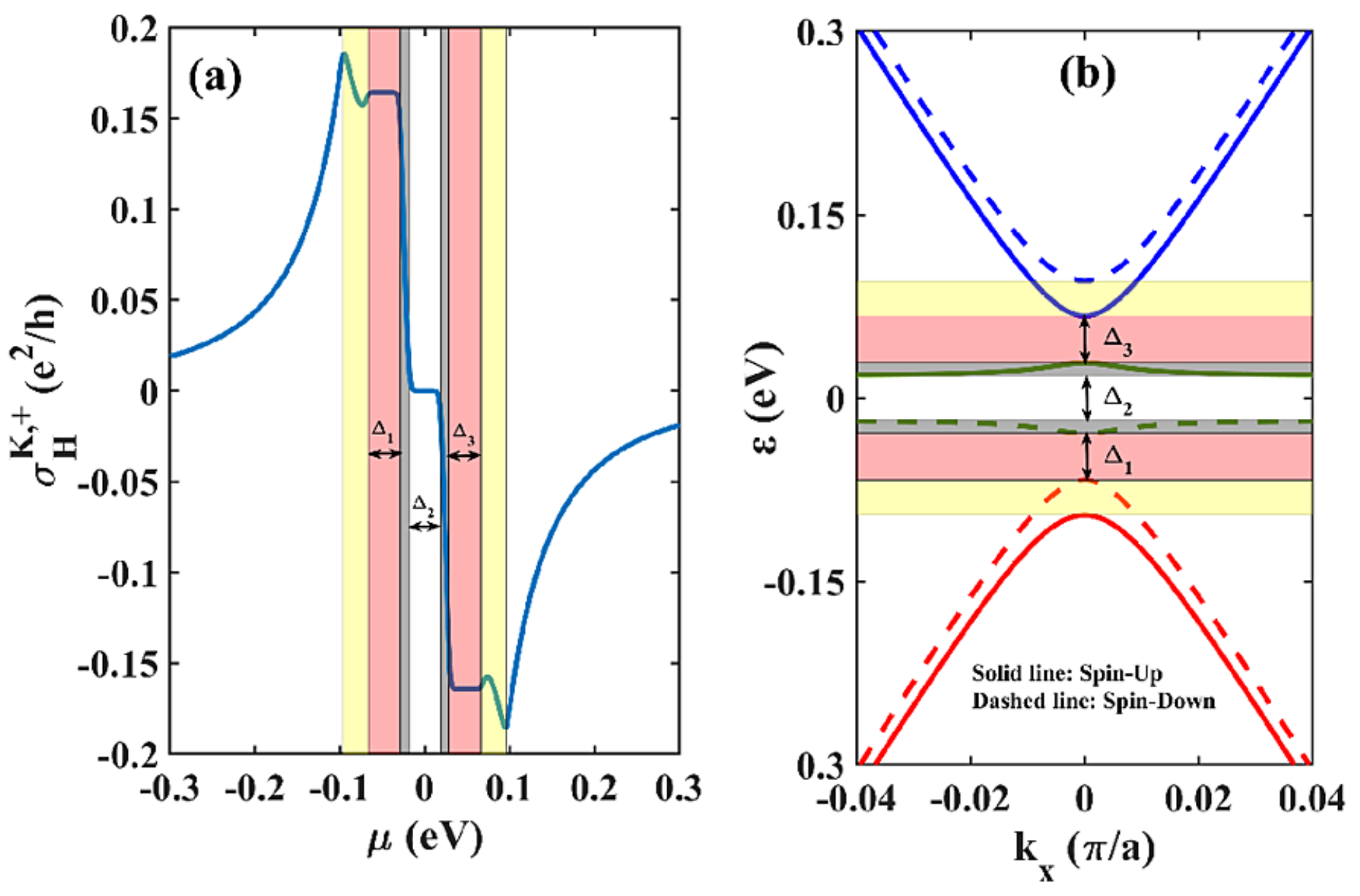}
\caption{(a) Spin summed AHC for the $K$ valley at $T = 20$~K and (b) Corresponding low-energy band structure for $\alpha=0.3$ highlighting the band gaps $\Delta_{1}=38$ meV, $\Delta_{2}= 40$ meV, and $\Delta_{3}=\Delta_1$.
Here, we consider $\lambda = 100$~meV and $M=0$.}
\label{fig:Spin_Valley_Hall2}
\end{figure}

To clarify the plateau structure in the VHC, we show the chemical potential dependence of the spin summed AHC in the $K$ valley at $\alpha=0.3$ in Fig.~\ref{fig:Spin_Valley_Hall2}(a). It is evident that $\sigma_{\rm H}^{K,+}$ attains a plateau whenever $\mu$ falls in one of SOI induced gaps ($\Delta_1$, $\Delta_2$, and $\Delta_3$) in the band structure, as shown in Fig.~\ref{fig:Spin_Valley_Hall2}(b). In the $T\to 0$ limit, the spin- and valley-resolved AHC given in Eq. \eqref{ahc} becomes
\begin{eqnarray}
\sigma_{\rm H}^{\eta,\sigma}=
\frac{e^2}{h} \sum_{n\in \rm occ}\int \frac{d^2\bm{k}}{2\pi}
\Omega_{\eta,\sigma}^n(\bm{k})=\frac{e^2}{h}\sum_{n\in \rm occ}C_{\eta,\sigma}^{(n)},
\end{eqnarray}
where $C_{\eta,\sigma}^{(n)}$ is the spin- and valley-resolved Chern number associated with the 
$n^{\rm th}$ occupied band. An explicit calculation yields~\cite{Orbit_M}
$C_{K,\uparrow}^{(2)}=0.581$, $C_{K,\downarrow}^{(2)}=-0.416$ and $C_{K,\uparrow}^{(1)}=C_{K,\downarrow}^{(1)}=-0.165$ for $\alpha=0.3$. The TRS further dictates that $C_{K,\uparrow}^{(n)}=-C_{K^\prime,\downarrow}^{(n)}$ and $C_{K,\downarrow}^{(n)}=-C_{K^\prime,\uparrow}^{(n)}$. We emphasize that $C^{(n)}_{\eta,\sigma}$ is evaluated within the low-energy continuum theory around valley $\eta$. Therefore, these quantities are not topological invariants defined over the full Brillouin zone and, consequently, are not required to be integer-valued. Nevertheless, they provide valuable insight into the low-energy contributions to the anomalous Hall plateaus. The integer-valued spin Chern number~\cite{Spin_Hall_Phase, Orbit_M} is recovered only after summing over all occupied bands as well as both spin and valley degrees of freedom:
\begin{eqnarray}
C_s=\frac{1}{2}\sum_{\substack{\eta,\, \sigma\\  n\in\text{occ}}}
\sigma\,C^{(n)}_{\eta,\sigma}.
\end{eqnarray} 

Now the behavior of the spin integrated AHC in the $K$ valley, i.e., $\sigma_{\rm H}^{K,+}$, can be understood from the following equation:
\begin{eqnarray}\label{Hall_occ}
\sigma_{\rm H}^{K,+}=\frac{e^2}{h}\sum_{n\in \rm occ}\big(C_{K,\uparrow}^{(n)}+C_{K,\downarrow}^{(n)}\big).
\end{eqnarray}
As $\mu$ scans the entire energy spectrum, 
$\sigma_{\rm H}^{K,+}$ increases monotonically until
$\mu$ reaches the top of the spin-up VB. This increase arises because the Berry curvature of the spin-up VB is positive for all $\bm k$-values, and the corresponding Chern number is also positive. 
As $\mu$ is increased further (shaded yellow region), $\sigma_{\rm H}^{K,+}$ begins to decrease since the spin-down VB starts contributing. The Berry curvature of this band is negative for most $\bm k$ values, which is reflected in its negative Chern number. When $\mu$ lies within the gap $\Delta_1$, $\sigma_{H}^{K,+}$ attains a plateau. In this regime, both spin-up and spin-down VBs are fully occupied. Consequently, $C_{K,\uparrow}^{(2)}+C_{K,\downarrow}^{(2)}=0.165$. According to 
Eq. \eqref{Hall_occ}, this yields $\sigma_{\rm H}^{K,+}=0.165\, e^2/h$. As $\mu$ enters a narrow shaded (grey) region, the spin-down FB begins to be occupied. Since its Berry curvature is predominantly negative over almost the entire $\bm k$ values (corresponding to a negative Chern number), 
$\sigma_{\rm H}^{K,+}$ decreases sharply and settles into another plateau once $\mu$ reaches the gap $\Delta_2$. When $\mu$ is in $\Delta_2$, the sub-bands $\varepsilon_{K,\uparrow}^{(2)}$, $\varepsilon_{K,\downarrow}^{(2)}$, and $\varepsilon_{K,\downarrow}^{(1)}$ are fully occupied. Hence, Eq. \eqref{Hall_occ} dictates that $\sigma_{\rm H}^{K,+}=0$, since $C_{K,\uparrow}^{(2)}+C_{K,\downarrow}^{(2)}+C_{K,\downarrow}^{(1)}=0$. Upon further increasing 
$\mu$, $\sigma_{\rm H}^{K,+}$ exhibits another sharp decrease as the spin-up FB starts to be populated. The Berry curvature of this band is identical to that of the spin-down FB, and consequently, it carries the same Chern number. When $\mu$ enters the gap  $\Delta_3=\Delta_1$, $\sigma_{\rm H}^{K,+}$ again attains a plateau. In this regime, all sub-bands of the VB and FB are fully occupied. The height of this plateau is solely determined by the Chern number of the spin-up FB, yielding $\sigma_{\rm H}^{K,+}=-0.165\, e^2/h$. As $\mu$ is increased further, a similar sequence repeats, with the CB playing the role previously taken by the VB. The behavior of the VHC [Fig.~\ref{fig:svh_salp}(a)] and 
SHC [Fig.~\ref{fig:svh_salp}(b)] can be understood along the same lines as that of 
$\sigma_{\rm H}^{K,+}$.

\begin{figure}[h!]
\centering
\includegraphics[width=8 cm, height=3.8 cm]{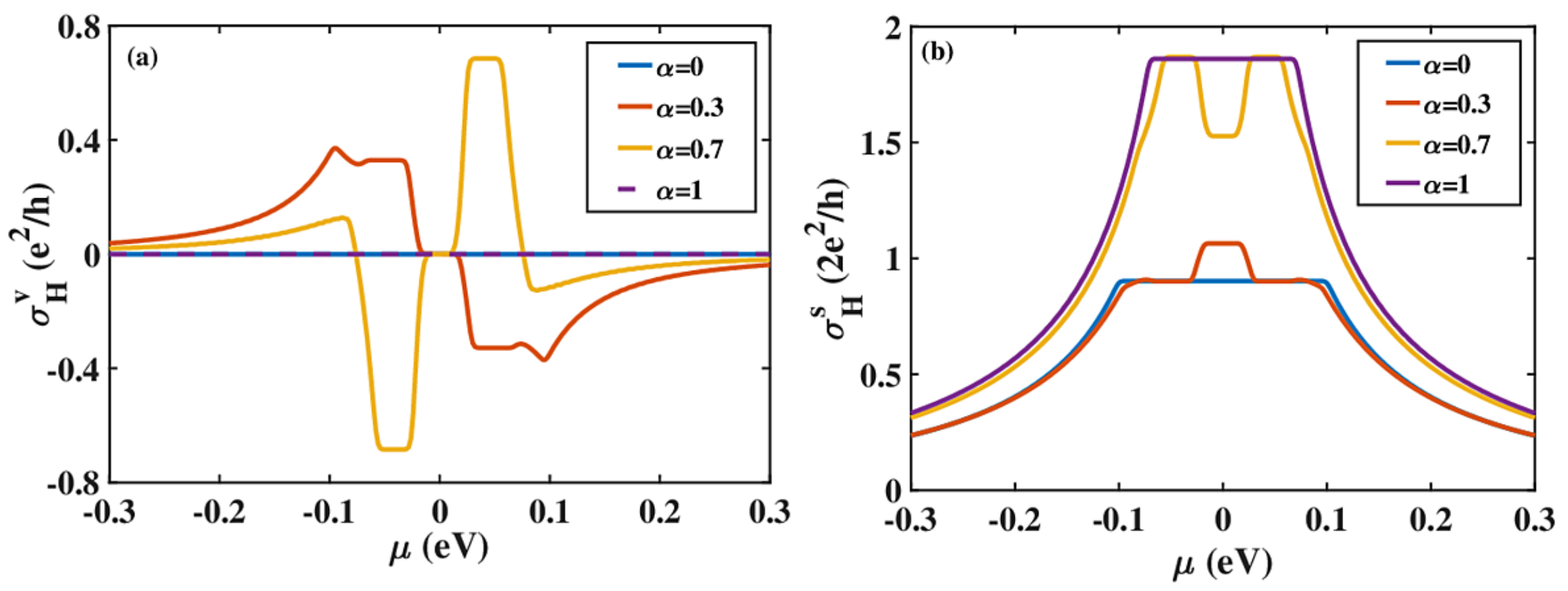}
\caption{(Color online) Dependence of the (a) Valley Hall conductivity and (b) Spin Hall conductivity on $\mu$ for several $\alpha$ values at $M=0$.}
\label{fig:SHC_alpha_Mzero}
\end{figure}

In Fig.~\ref{fig:SHC_alpha_Mzero}(a) and \ref{fig:SHC_alpha_Mzero}(b), we show the chemical potential dependence of the VHC and SHC, respectively, for different values of $\alpha$. As evident from Fig.~\ref{fig:SHC_alpha_Mzero}(a), the VHC vanishes for $\alpha=0$ and $\alpha=1$, as a combined effect of TRS and IS.
For an intermediate value of $\alpha$ ($0 < \alpha < 1$), a finite valley Hall response emerges as a consequence of broken IS. The behavior of the VHC for $\alpha = 0.3$ has already been discussed in the previous paragraph. For $\alpha = 0.7$, the VHC exhibits similar plateau-like characteristics to those observed for $\alpha = 0.3$, and thus the overall explanation remains the same. However, a striking difference is observed: the plateaus appearing in the gap $\Delta_1$ and $\Delta_3$ have opposite signs for $\alpha = 0.3$ and $\alpha = 0.7$. More specifically, the heights of the plateau in both $\Delta_1$ and $\Delta_3$ undergo a sign change across $\alpha = 0.5$, indicating the occurrence of a TPT, as confirmed in recent studies~\cite{Spin_Hall_Phase, Orbit_M}. Furthermore, the widths of the plateaus vary for different values of $\alpha$ because the spin–orbit gaps are $\alpha$-dependent.

On the other hand, the SHC for $\alpha=0$ ($\alpha=1$) exhibits a smooth plateau $\sigma_{\rm H}^{\rm s}\sim 2e^2/h$ ($4e^2/h$), when $\mu$ lies  within the band gap. 
For $\alpha = 0$, the FB is absent, and only a single energy gap exists between the CB and the VB, which gives rise to the quantized plateau.
For $\alpha = 1$, although the FB is present, it remains undistorted by the SOI. In this case, two gaps appear in the spectrum: one between the CB and FB and another between the FB and VB. These gaps are not separated but share a common boundary through the FB. The presence of these gap regions in the spectrum leads to the wide and smooth SHC plateau observed in these limits.
The SHC exhibits three distinct plateaus for $0<\alpha<1$, as the SOI opens three well-separated energy gaps, $\Delta_1$, $\Delta_2$, and $\Delta_3$. The plateau height in $\Delta_1$ (as well as in $\Delta_3$) undergoes an abrupt jump from $2e^2/h$ to $4e^2/h$ across $\alpha = 0.5$, consistent with the corresponding change in the spin Chern number from $C_s = 1$ to $C_s = 2$, thereby confirming the occurrence of a TPT.

\subsubsection{Case II: Hall Conductivities in the Presence of Magnetization ($M \neq 0$)}
Here, we discuss the characteristics of the Hall conductivities in the presence of a staggered magnetization $M\neq0$. As observed earlier, the magnetization modifies the magnitudes and positions of the energy gaps. It also redistributes the Berry curvature around the valleys, thereby significantly influencing the transport responses.

\begin{figure}[H]
\centering
\includegraphics[width=8 cm, height=3.8 cm]{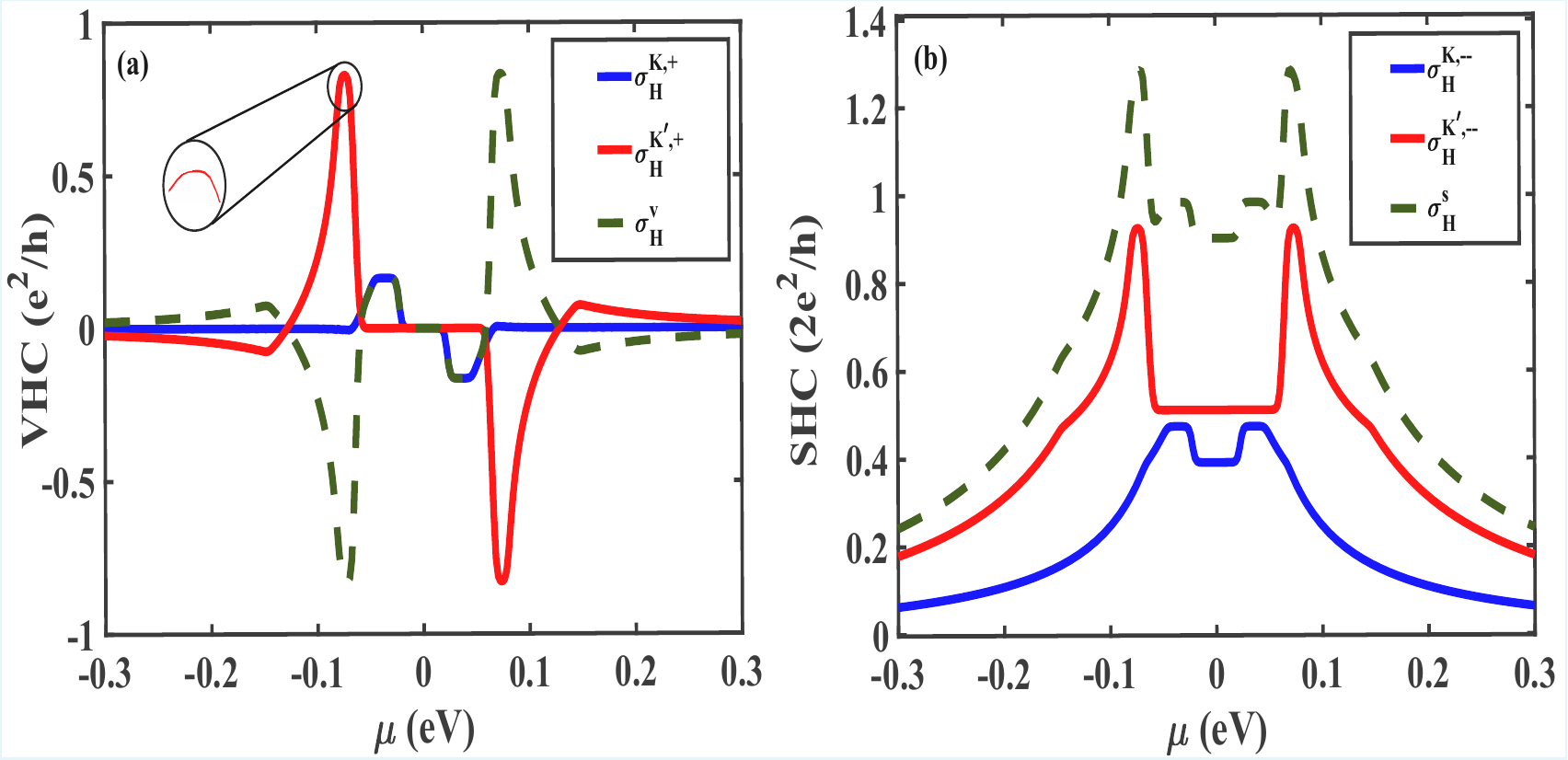}
\caption{(Color online) (a) Valley Hall conductivity and (b) Spin Hall conductivity for $\alpha=0.3$ as a function of chemical potential $\mu$ with $M=50$ meV. The plateau in $\sigma_{\rm H}^{K^\prime,+}$, which apparently appears as a peak, is shown in the zoomed portion.}
\label{fig:svh_dalp}
\end{figure}

Figures~\ref{fig:svh_dalp}(a) and \ref{fig:svh_dalp}(b) show the VHC and SHC as a function of the chemical potential $\mu$ for $\alpha = 0.3$ and $M = 50~\mathrm{meV}$. Due to the breaking of the TRS, the $K$ and $K^\prime$ valleys contribute differently to the VHC and SHC. In particular, their chemical potential dependences become markedly distinct. While the behavior of $\sigma_{\rm H}^{K,+}$ remains similar to the corresponding case with $M = 0$, $\sigma_{\rm H}^{K^\prime,+}$ deviates significantly from its nonmagnetic counterpart. A similar distinction is also present between $\sigma_{\rm H}^{K,-}$ and $\sigma_{\rm H}^{K^\prime,-}$. According to Fig. \ref{fig:Energy}(c) and \ref{fig:Energy}(d) the gap between spin-up VB and spin-up FB in the $K^\prime$ valley becomes much narrower compared to that in the $K$ valley. A similar reduction occurs for the gap between the spin-down CB and spin-down FB. In contrast, the gap between the spin-up FB and spin-down FB in the $K^\prime$ valley becomes significantly wider than that in the $K$ valley, leading to the appearance of a wide central plateau in $\sigma_{\rm H}^{K^\prime,+(-)}$. In addition, the positions of the band gaps are shifted between the two valleys, which causes the plateaus to occur over different ranges of $\mu$ in the $K$ and $K^\prime$ valleys. Overall, these features explain the plateau structures observed in 
$\sigma_{\rm H}^{\rm v}$ and $\sigma_{\rm H}^{\rm s}$. Notably, a finite anomalous Hall response persists in addition to the spin and valley Hall responses. 

\begin{figure}[h!]
\centering
\includegraphics[width=8 cm, height=3.8 cm]{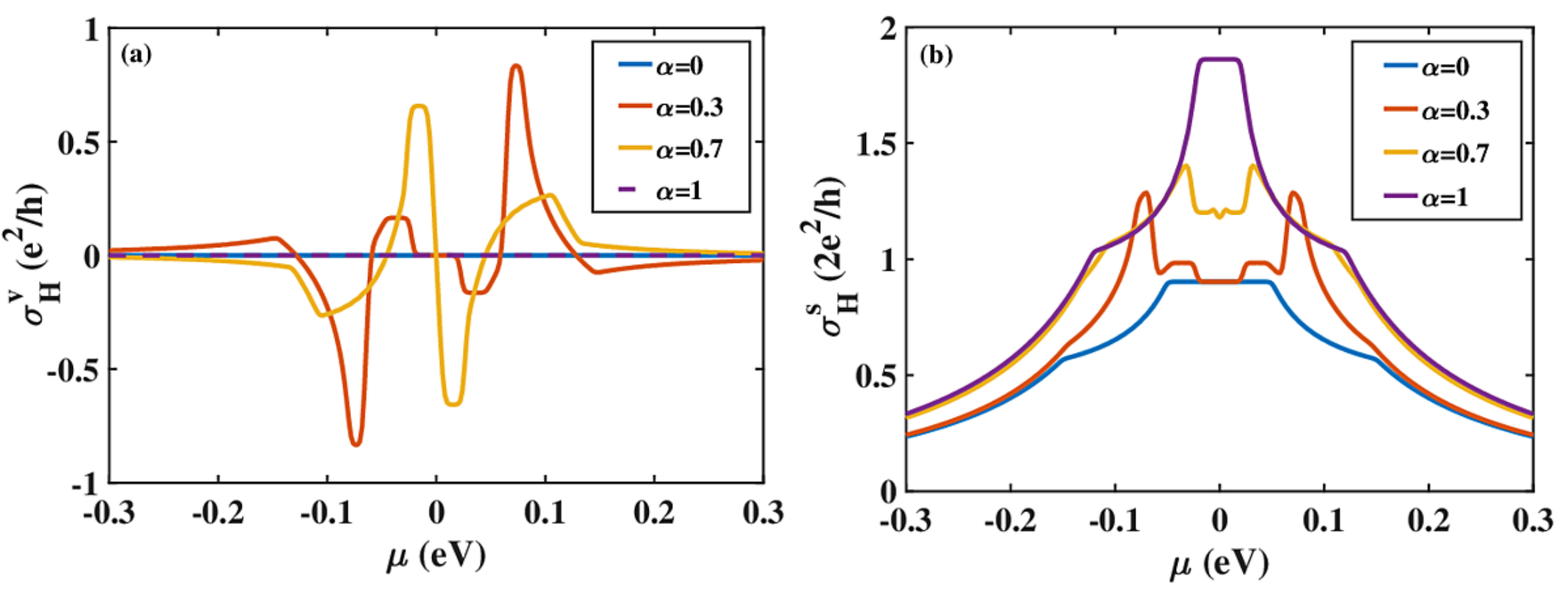}
\caption{(Color online) Dependence of the (a) Valley Hall conductivity and (b) Spin Hall conductivity on $\mu$ for several $\alpha$ values at $M=50$ meV.}
\label{fig:shc_salp}
\end{figure}

Figures ~\ref{fig:shc_salp}(a) and \ref{fig:shc_salp}(b) illustrate the $\mu$-dependence of the VHC and SHC, respectively, for different $\alpha$. For $\alpha = 0$ and $\alpha = 1$, IS is preserved, and hence the VHC vanishes despite the breaking of TRS. However, for 
$0 < \alpha < 1$, broken IS leads to a finite VHC, which differs significantly from that of the corresponding nonmagnetic case, both in character and magnitude. On the other hand, the SHC for 
$\alpha=0$ and $\alpha=1$ exhibit quantized values $2e^2/h$ and $4e^2/h$, respectively, when the chemical potential lies in the energy gap, similar to the $M=0$ case. This indicates that the system remains in QSH phases with $C_s=1$ and $C_s=2$ respectively, even in the presence of finite $M$. However, this is not the case for $0<\alpha<1$, where the interplay between $M$ and $\alpha$ gives rise to a rich topological phase diagram~\cite{Spin_Hall_Phase}, as mentioned earlier. For fixed values of $M$ and $\lambda$, two QSH phases with $C_s=1$ and $C_s=2$ are separated by a Quantum spin quantum anomalous Hall phase (QSQAH) with $C=1$ and $C_s=3/2$ as established in Ref. \cite{Spin_Hall_Phase}. For a given $M<\lambda/\sqrt{2}$, the phase boundaries are determined by the critical values of 
$\alpha$ derived in Ref. \cite{Orbit_M}
\begin{eqnarray}\label{alph_bn}
\alpha_\pm=\frac{2\pm\sqrt{5m^2-m^4}}{4-m^2},
\end{eqnarray}
where $m=M/\lambda$. For the chosen parameters, $M=50$ meV and $\lambda=100$ meV, i.e, $m=1/2$, we find that  Eq.~\eqref{alph_bn} agrees well with our numerical calculation yielding, the critical values $\alpha_-\approx0.24$ and $\alpha_+\approx0.82$. Therefore, the considered values of $\alpha$ in Figs. \ref{fig:shc_salp}(a) and \ref{fig:shc_salp}(b), i.e., $\alpha=0.3$ and $\alpha=0.7$ correspond to the QSQAH phase. Consequently, the corresponding spin and valley Hall responses differ significantly from those of the pure QSH phases.

In the $T\rightarrow0$ limits, it is possible to determine the analytical expression for the AHC in the $\alpha=0$ and $\alpha=1$ limits. When the chemical potential $\mu$ lies in the VB(CB), we find for $\alpha=1$
\begin{equation}\label{AHC_dice}
    \sigma_{\rm H}^{\eta,\sigma} =+(-)\eta \frac{e^2}{h}\Bigg[\text{sgn}(\Delta_{\rm d})-\frac{\Delta_{\rm d}}{\sqrt{(\mu^2+\Delta_{\rm d}^2)}}\Bigg].
\end{equation}
Similarly, for the $\alpha=0$ case, the AHC is obtained as

\begin{equation}\label{AHC_graphene}
    \sigma_{\rm H}^{\eta,\sigma} =+(-)\eta \frac{e^2}{2h}\Bigg[\text{sgn}(\Delta_{\rm g})-\frac{\Delta_{\rm g}}{\sqrt{(\mu^2+\Delta_{\rm g}^2)}}\Bigg].
\end{equation}
Using Eqs. (\ref{AHC_dice}) and (\ref{AHC_graphene}), it is straightforward to show that the VHC vanishes in the limiting cases, i.e., $\alpha=0$ and $\alpha=1$.

\subsection{Valley and spin Nernst conductivities}

In this section, we present a thorough analysis of the Nernst responses of the system in the absence and presence of the staggered magnetization. The VNC and SNC are calculated numerically using Eqs.~(\ref{eq:anc}) and~(\ref{SVNC}).  In the following, we discuss the characteristic features of the VNC and SNC separately in the two physically distinct regimes: the time-reversal–symmetric case ($M=0$) and the time-reversal–broken case ($M\neq 0$). We set temperature $T=20$ K for all cases  in this section.

\subsubsection{Case I: Nernst Conductivities in the absence of Magnetization ($M = 0$)}

\begin{figure}[h!]
\centering
\includegraphics[width=8.5 cm, height=4 cm]{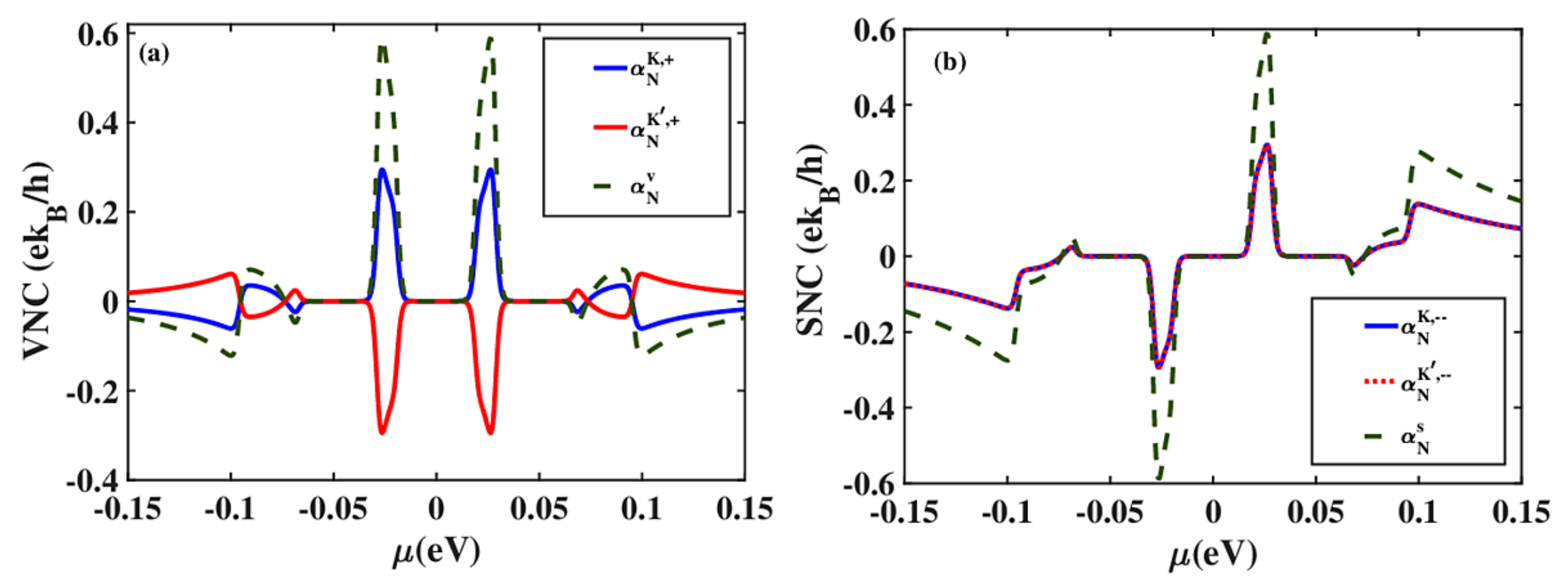}
\caption{(Color online) (a) Valley Nernst conductivity and (b) Spin Nernst conductivity as a function of chemical potential $\mu$ for $\alpha=0.3$ and $M=0$.}
\label{fig: vnc_salp}
\end{figure}

Figure~\ref{fig: vnc_salp}(a) depicts the chemical potential dependence of the spin-summed ANC in individual valleys along with the VNC, while Fig.~\ref{fig: vnc_salp}(b) shows the SNC at individual valleys and the total SNC, considering 
$\alpha=0.3$. The VNC and SNC are explicitly given by $\alpha_{\rm N}^{\rm v}=\alpha_{\rm N}^{K,+}-\alpha_{\rm N}^{K^\prime,+}$ and $\alpha_{\rm N}^{\rm s}=\alpha_{\rm N}^{K,-}+\alpha_{\rm N}^{K^\prime,-}$, respectively, where $\alpha_{\rm N}^{\eta,\pm}=\alpha_{\rm N}^{\eta,\uparrow}\pm\alpha_{\rm N}^{\eta,\downarrow}$ for $\eta=K,K^\prime$.
The VNC contributions from the $K$ and $K^{\prime}$ valleys are equal in magnitude but opposite in sign due to the TRS, resulting in a vanishing total ANC (both spin and valley integrated). Nevertheless, a finite VNC would persist as shown by the dashed line in Fig.~\ref{fig: vnc_salp} (a). The VNC also appears to be a symmetric function of $\mu$, unlike the VHC. On the other hand, the SNC is valley-degenerate. Therefore, the total SNC (summed over the valley indices) is doubled. In contrast to the VNC, the SNC varies antisymmetrically with the chemical potential. Despite their key differences, both the VNC and SNC exhibit similar features, namely pronounced peak–dip structures and intermediate zero-value plateaus. 

\begin{figure}[h!]
\centering
\includegraphics[width=8.2 cm, height=4 cm]{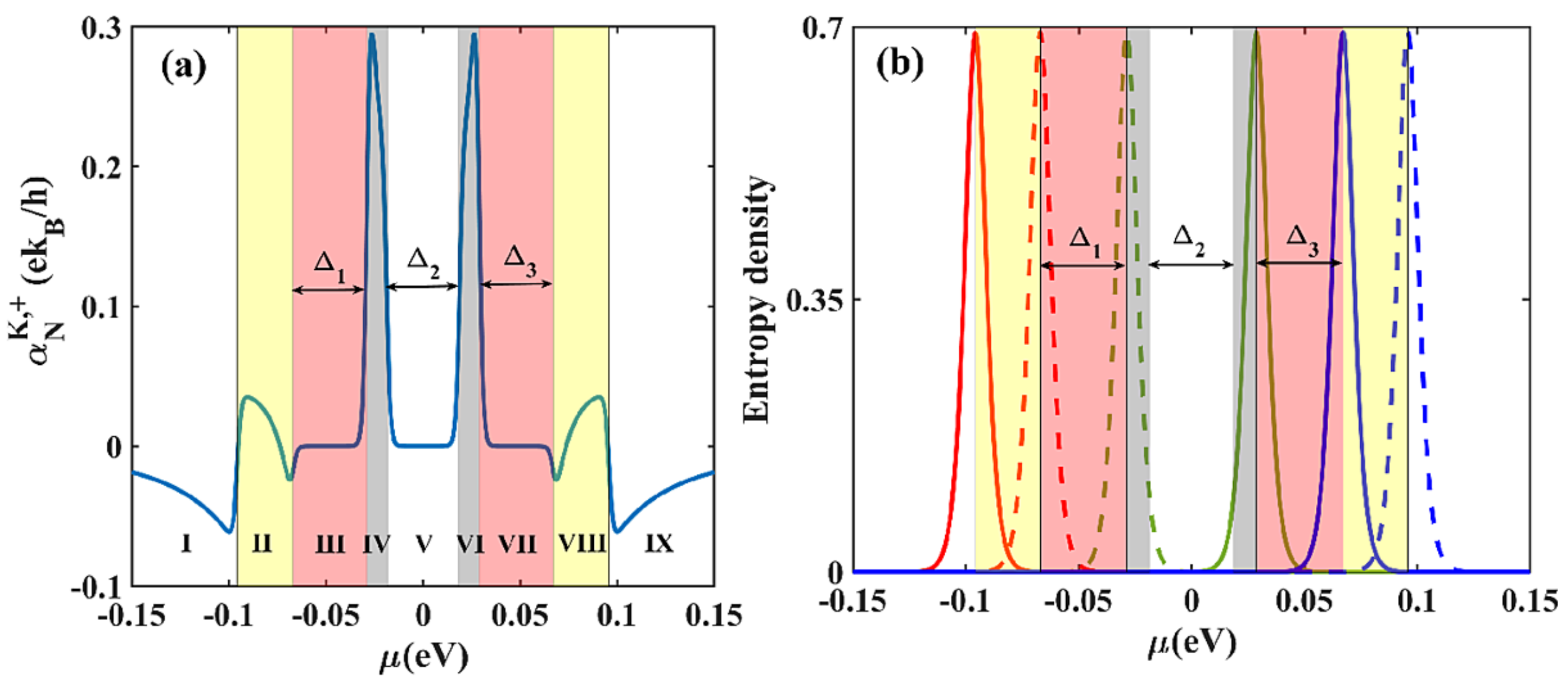}
\caption{(Color online) 
(a) Spin summed ANC in the $K$ valley and (b) corresponding entropy density for 
$\alpha=0.3$.  
Sharp peak–dip features in panel (a) coincide with the entropy maxima in panel (b) and 
occur near band edges, where the Berry curvature is strongly enhanced.  
The shaded regions III, V, and VII denote the band gaps in which both entropy and $\alpha_{\rm N}^{K,+}$ are suppressed.}
\label{fig:vnc_entropy}
\end{figure}

To understand the peak–dip features of the VNC and SNC, we replot $\alpha_{\rm N}^{K,+}$ in Fig.~\ref{fig:vnc_entropy}(a), with the various regions of the spectrum indicated by shaded areas. The zero-value plateau regions of $\alpha_{\rm N}^{K,+}$ correspond to the SOI-induced band gaps 
$\Delta_1$, $\Delta_2$, and $\Delta_3$.
Generally, the characteristic features of the ANC are mainly governed by the interplay between the Berry curvature and the entropy density. The entropy density function $S^{n}_{\eta,\sigma}(\mathbf{k})$, which enters directly in Eq.~(\ref{eq:anc}), depends solely on the Fermi-Dirac distribution function $f$. At low temperatures, it shows a typical step-like profile and vanishes for fully occupied ($f=1$) or empty states ($f=0$). At half filling ($f=1/2$), it reaches a maximal value of $\ln 2$, evident from Eq. \eqref{eq:entropy}. The variation of the entropy density with $\mu$ is shown in Fig.~\ref{fig:vnc_entropy}(b). When $\mu$ approaches a band edge, the entropy density becomes sharply peaked within an energy window of the order of a few $k_BT$. The overlap of the entropy density with the Berry curvature hotspots near band extrema provides the key mechanism responsible for the peak-dip structures in $\alpha_{\rm N}^{K,+}$. 
For instance, $\alpha_{\rm N}^{K,+}$ decreases monotonically as $\mu$ approaches the top of the spin-up VB [Region I in Fig.\ref{fig:vnc_entropy} (a)]. This occurs because the entropy density increases and the Berry curvature is positive for all $\bm{k}$, with a peak at $\bm{k}=0$. As $\mu$ crosses the spin-up VB $\alpha_{\rm N}^{K,+}$ undergoes a sign change from $-$ to $+$. At this point, states from the spin-down VB begin to contribute, whose Berry curvature is negative over certain regions of $\bm{k}$ space and dominates over the contribution from the spin-up VB. Consequently, $\alpha_{\rm N}^{K,+}$ decreases within the shaded (yellow) region II and attains a dip when $\mu$ reaches the top of the spin-down VB. This behavior results from the combined effect of the Berry curvature of the VBs and the entropy density.
When $\mu$ lies within the band gap $\Delta_1$ (region III), $\alpha_{\rm N}^{K,+}$ vanishes because the entropy density becomes zero in the gap, as evident from Fig.~\ref{fig:vnc_entropy}(b). The small deviation from zero near the edges of $\Delta_1$ arises from finite-temperature effects. As $\mu$ enters the shaded gray region IV, $\alpha_{\rm N}^{K,+}$ exhibits a large peak whose width is of the order of that region. In this regime, the states of the spin-down FB become occupied; their Berry curvature is negative for all $\bm{k}$ and exhibits a pronounced dip at $\bm{k}=0$. Although the entropy density decreases across this region, it remains positive, which leads to the observed peak in $\alpha_{\rm N}^{K,+}$.
When $\mu$ enters the gap $\Delta_2$ (region V), $\alpha_{\rm N}^{K,+}$ again exhibits a zero-value plateau due to the vanishing entropy density $S_{\eta,\sigma}^{n}$. 
For $\mu>0$ (regions VI-IX), the same behavior repeats with the role of the VBs replaced by that of the CBs. The explanations of the peak-dip structures in the VNC and SNC follow the same reasoning as in the case of 
$\alpha_{\rm N}^{K,+}$.

Within the considered temperature regime, where $k_BT<<\vert \mu\vert$, it is expected that the behavior of the ANC shown in Fig.~\ref{fig: vnc_salp} can also be understood from the Mott relation, since phonon-drag effects, strong electron correlations, and inelastic scattering processes are not considered. In the low temperature limit, it is straightforward to derive the following Mott relation from Eq.~\eqref{anc_alt}
\begin{eqnarray}
\alpha_{\rm N}^{\eta,\sigma} = -\frac{\pi^{2}k_{B}^{2}T}{3e}\left.\frac{d\sigma_{H}^{\eta,\sigma}}{d\mu}\right|_{\mu=\varepsilon_F},   
\end{eqnarray}
where $\varepsilon_F$ is the Fermi energy.
Comparing Figs.~\ref{fig:Spin_Valley_Hall2}(a) and \ref{fig:vnc_entropy}(a), it is clear that $\alpha_{\rm N}^{K,+}$ is proportional to the negative derivative of $\sigma_{\rm H}^{K,+}$ with respect to the chemical potential $\mu$, which naturally explains the entire structure of $\alpha_{\rm N}^{K,+}$, including the peak–dip features, plateau regions, and the sign changes near the band edges, as discussed in the following. 
In regions where $\sigma_{\rm H}^{K,+}$ is nearly constant as a function of 
$\mu$, its derivative is close to zero, resulting in the plateau-like behavior of $\alpha_{\rm N}^{K,+}$. When $\sigma_{\rm H}^{K,+}$ varies rapidly with $\mu$, the magnitude of its derivative becomes large, leading to pronounced peak–dip structures in $\alpha_{\rm N}^{K,+}$. Because of the negative proportionality, the sign of $\alpha_{\rm N}^{K,+}$ is opposite to the slope of $\sigma_{\rm H}^{K,+}$ that means a positive(negative) slope of $\sigma_{\rm H}^{K,+}$ produces a negative(positive) contribution to $\alpha_{\rm N}^{K,+}$. Accordingly, at the band edges, where the slope of $\sigma_{\rm H}^{K,+}$ reverses, $\alpha_{\rm N}^{K,+}$ undergoes a sign change.

\begin{figure}[h!]
\centering
\includegraphics[width=8 cm, height=3.8 cm]{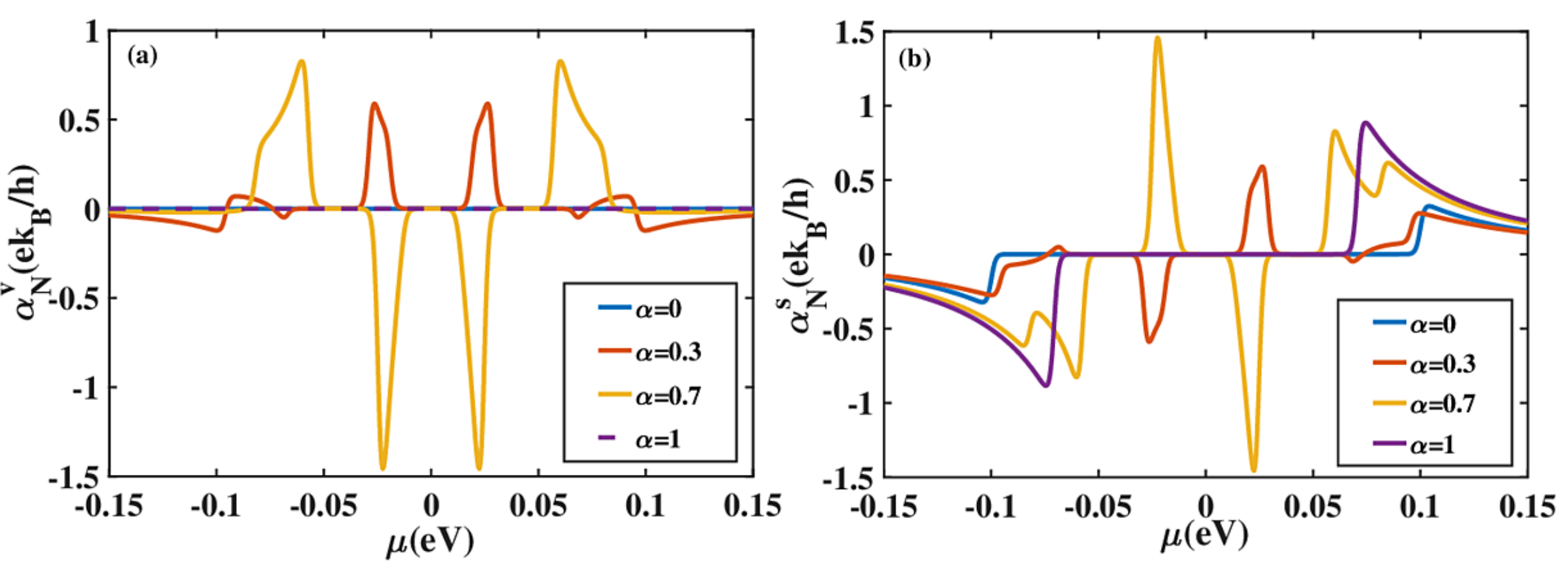}
\caption{(Color online) Dependence of the (a) Valley Nernst conductivity and (b) Spin Nernst conductivity on $\mu$ for several $\alpha$ values at $M=0$.}
\label{fig: vnc_dalp}
\end{figure}

The variation of the VNC and SNC with 
$\mu$ for different values of $\alpha$ are shown in Figs.~\ref{fig: vnc_dalp}(a) and \ref{fig: vnc_dalp}(b), respectively. The simultaneous presence of TRS and IS enforces a vanishing VNC for $\alpha=0$ and $\alpha=1$, due to $S^{n}_{\eta,\uparrow} = S^{n}_{\eta,\downarrow}$ and $\Omega^{n}_{\eta,\uparrow} = -\Omega^{n}_{\eta,\downarrow}$. A finite VNC appears for $0 < \alpha < 1$ owing to broken IS. The peak-dip features of VNC for $\alpha=0.3$ and $\alpha=0.7$ are completely different. It is verified (not shown here) that all the major peaks/dips undergo a sign change across $\alpha=0.5$. This may be considered as an indication of TPT. On the other hand, the SNC exhibits wide zero-value plateaus for $\alpha=0$ and $\alpha=1$. Similar to the VNC, the peaks and dips in the SNC for $0<\alpha<1$ would also display a sign change across $\alpha=0.5$.

\subsubsection{Nernst Conductivities in the Presence of Magnetization ($M \neq 0$)}
\begin{figure}[h!]
\centering
\includegraphics[width=8 cm, height=3.8 cm]{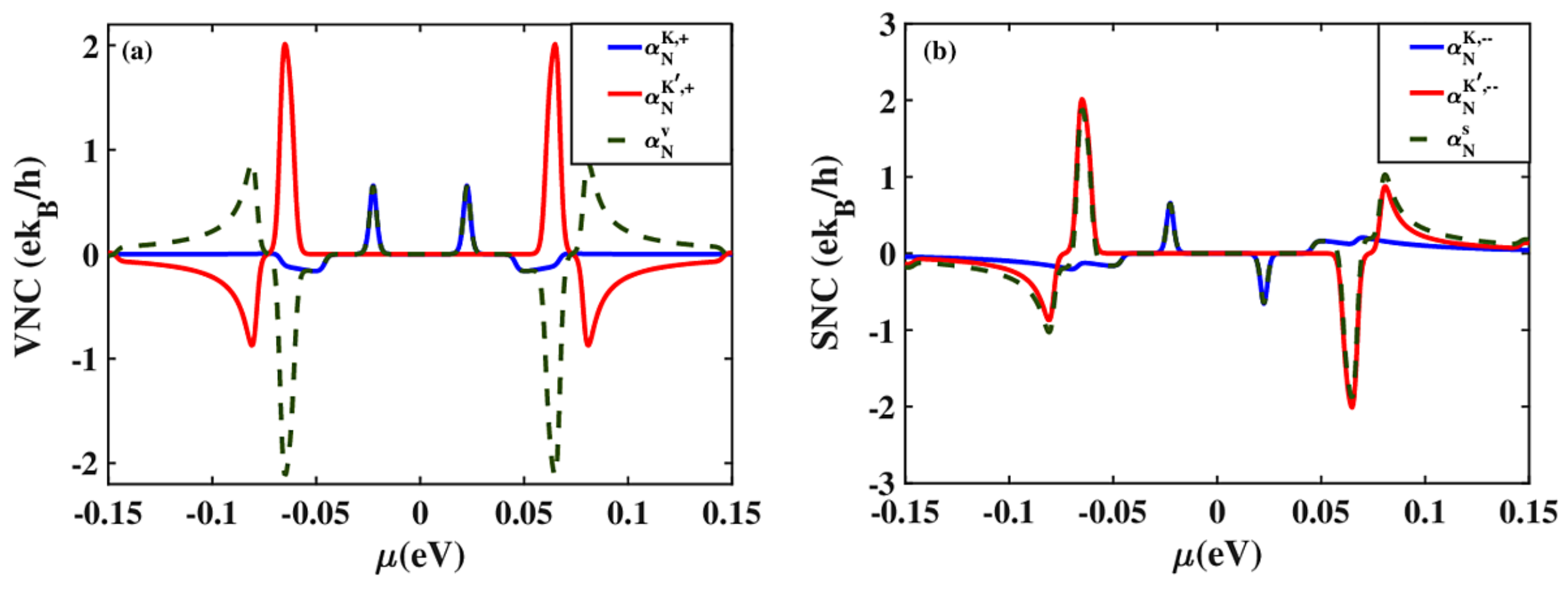}
\caption{(Color online) (a) Valley Nernst conductivity and (b) Spin Nernst conductivity  as a function of chemical potential $\mu$ with $\alpha=0.3$ and $M= 50$ meV.}
\label{fig: sns_salp}
\end{figure}

The presence of a staggered magnetization leads to a drastic change in the Nernst response of the system as shown in Fig. \ref{fig: sns_salp}. As shown in Fig. \ref{fig: sns_salp}, $K$ and $K^\prime$ valleys contribute differently due to breaking of the TRS. The peak-dip feature of $\alpha_{\rm N}^{K,+}$ is more prominent than that of $\alpha_{\rm N}^{K^\prime,+}$, thus leading to a nonvanishing total ANC (not shown) in addition to the finite VNC (dashed line). 
Figure \ref{fig: sns_salp}(b) shows that SNCs at individual valleys are completely different. In fact the total SNC is dominated by the contribution from the $K^\prime$ valley. 

\begin{figure}[h!]
\centering
\includegraphics[width=8 cm, height=3.8 cm]{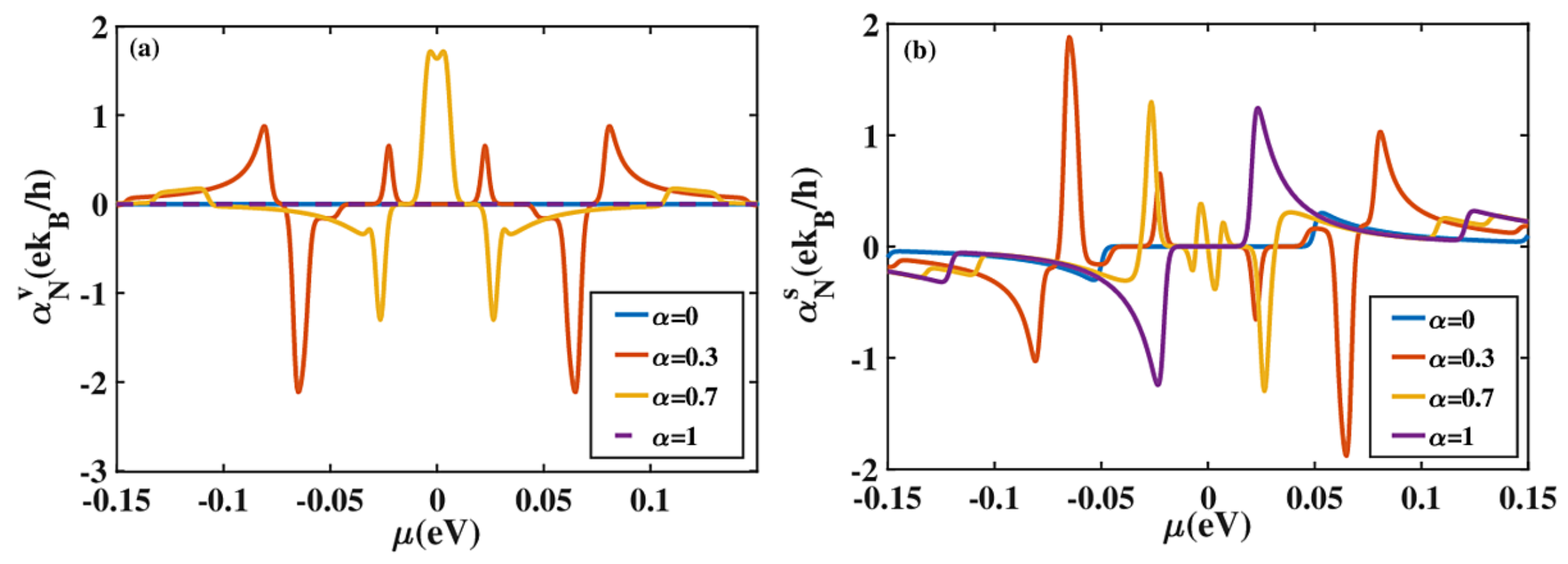}
\caption{(Color online) Dependence of the (a) Valley Nernst conductivity and (b) Spin Nernst conductivity on $\mu$ for several $\alpha$ values considering $M=50$ meV.}
\label{fig: snc_dalp}
\end{figure}

The dependence of the VNC and SNC on the chemical potential for various values of 
$\alpha$ are depicted in Figs.~\ref{fig: snc_dalp}(a) and \ref{fig: snc_dalp}(b), respectively. The VNC disappears for 
$\alpha=0$ and $\alpha=1$ due to IS. In contrast, the SNC remains finite and exhibits characteristic features in these limits. The other values of $\alpha$, namely $\alpha=0.3$ and $\alpha=0.7$ considered in Figs.~\ref{fig: snc_dalp}(a) and \ref{fig: snc_dalp}(b) correspond to the QSQSH phase with $C=1$ and $C_s=3/2$ for the chosen value of $M$. Figs.~\ref{fig: snc_dalp}(a) and \ref{fig: snc_dalp}(b) further confirm that the spin-valley Nernst responses in this mixed phase strongly depend on $\alpha$.

At low temperatures, using Eq. (\ref{eq:anc}), we obtain analytical expressions for the ANC in the $\alpha=0$ and $\alpha=1$ limits. The ANC takes a non-zero value when the chemical potential $\mu$ lies within the band and vanishes otherwise.
For $\alpha=1$, when the chemical potential $\mu$ lies in the CB(VB), we obtain the simpler form of the ANC as
\begin{equation}
    \alpha_{\rm N}^{\eta,\sigma}=+(-)\eta\frac{\pi k_B e T\Delta_{\rm d}}{6h\mu^2}
\end{equation}
and similarly for $\alpha=0$, the ANC is obtained as
\begin{equation}
    \alpha_{\rm N}^{\eta,\sigma}=+(-)\eta\frac{\pi k_B e T\Delta_{\rm g}}{12h\mu^2}.
\end{equation}

\subsection{Spin and Valley Polarization}
\begin{figure}[h!]
\centering
\includegraphics[width=8 cm, height=3.8 cm]{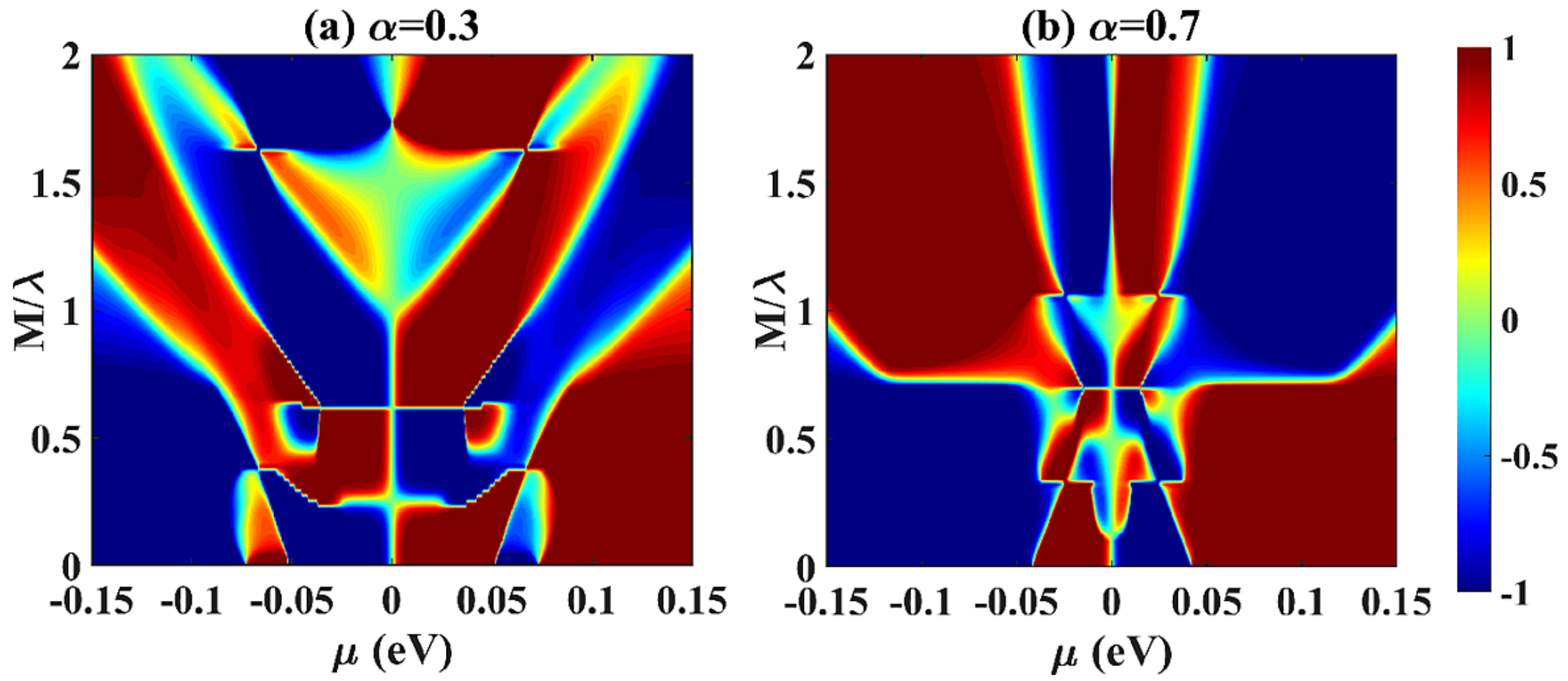}
\caption{(Color online) Contour plot of the spin polarization for (a) $\alpha=0.3$ and (b) $\alpha=0.7$.}
\label{fig: spin_pol}
\end{figure}

\begin{figure}[t]
\centering
\includegraphics[width=8 cm, height=3.8 cm]{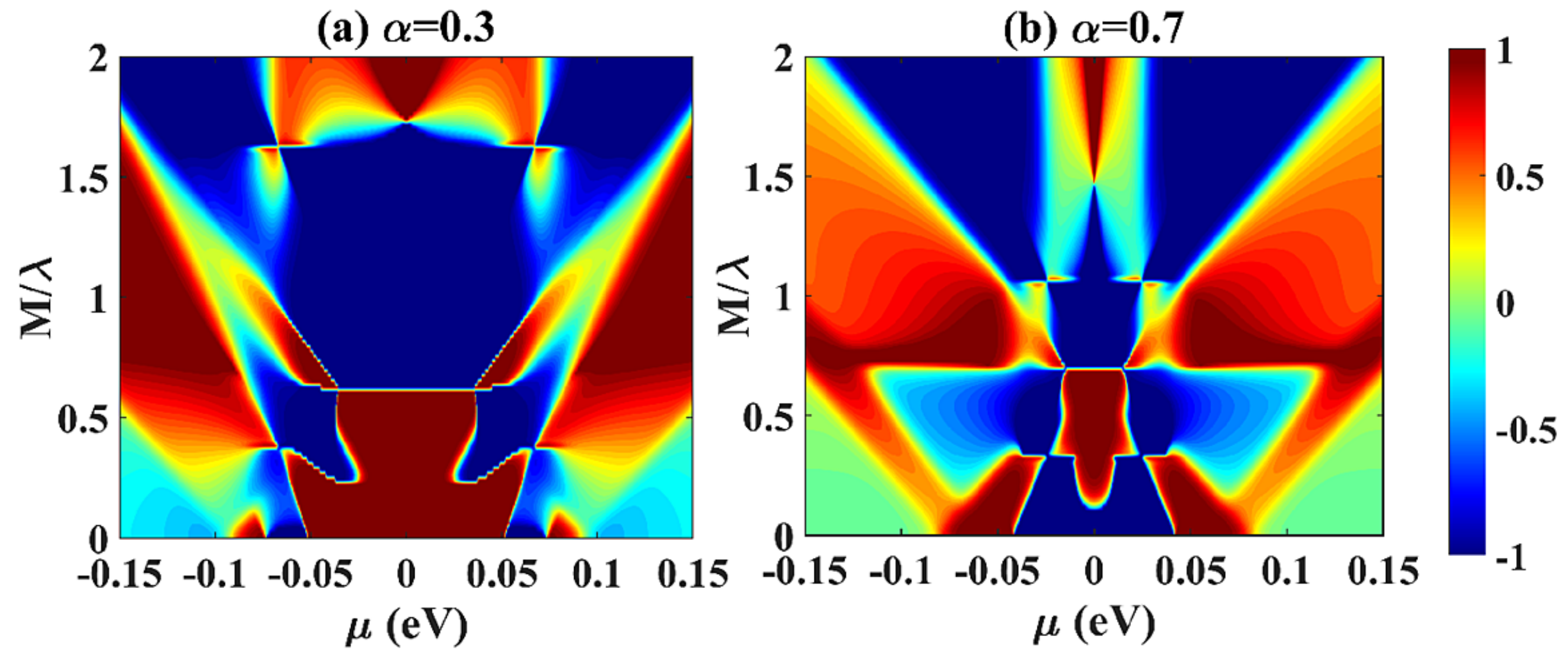}
\caption{(Color online)  Contour plot of the valley polarization for (a) $\alpha=0.3$ and (b) $\alpha=0.7$.}
\label{fig: valley_pol}
\end{figure}

In this section, we analyze the spin and valley polarization calculated from the Nernst conductivities in the presence of a staggered magnetization.   
The contour plot of the spin polarization $P_{\rm s}$ for $\alpha=0.3$ is shown in 
Fig.~\ref{fig: spin_pol}(a). The spin polarization exhibits distinct positive and negative regions in the $(\mu,M/\lambda)$, exhibiting clear asymmetry about $\mu=0$. Significant regions with nearly saturated polarization, $P_{\rm s}\approx \pm 1$, emerge over a wide range of chemical potential and magnetization, indicating strong spin-selective thermoelectric transport. The corresponding result for $\alpha=0.7$ is illustrated in Fig.~\ref{fig: spin_pol}(b). In this case, the spin polarization develops more sharply defined domains with a clear separation between the positive and negative regions. Broad regions with nearly complete spin polarization are observed throughout the $(\mu,M/\lambda)$ plane. The sharp interfaces between oppositely polarized states near $\mu=0$ signify an abrupt reversal of the dominant spin contribution under small variations of chemical potential or magnetization. 
Figure~\ref{fig: valley_pol}(a) shows the contour plot of the valley polarization $P_{\rm v}$ for $\alpha=0.3$. The valley polarization exhibits a symmetric distribution about $\mu=0$ with distinct regions of positive and negative polarization in the $(\mu,M/\lambda)$ plane. A prominent negatively polarized region with $P_{\rm v}\approx -1$ emerges in the central area of the $(\mu,M/\lambda)$ plane for intermediate values of $M/\lambda$, the other regions, however, are dominated by positive polarization. 
The corresponding result for $\alpha=0.7$ is shown in Fig.~\ref{fig: valley_pol}(b). Here, the valley polarization develops well-defined domains with sharper interfaces between the positive and negative regions. An extended region of positive polarization is observed for larger values of $|\mu|$. Additionally, a broad region with $P_{\rm v}\approx -1$ appears near $\mu=0$ and extends across a wide range of $M/\lambda$. Overall, the results demonstrate that the combined action of the staggered magnetization, spin-orbit interaction, and the lattice parameter ($\alpha$) provides an effective mechanism for controlling spin and valley-selective thermoelectric transport.

\begin{figure}[h!]
\centering
\includegraphics[width=8 cm, height=3.8 cm]{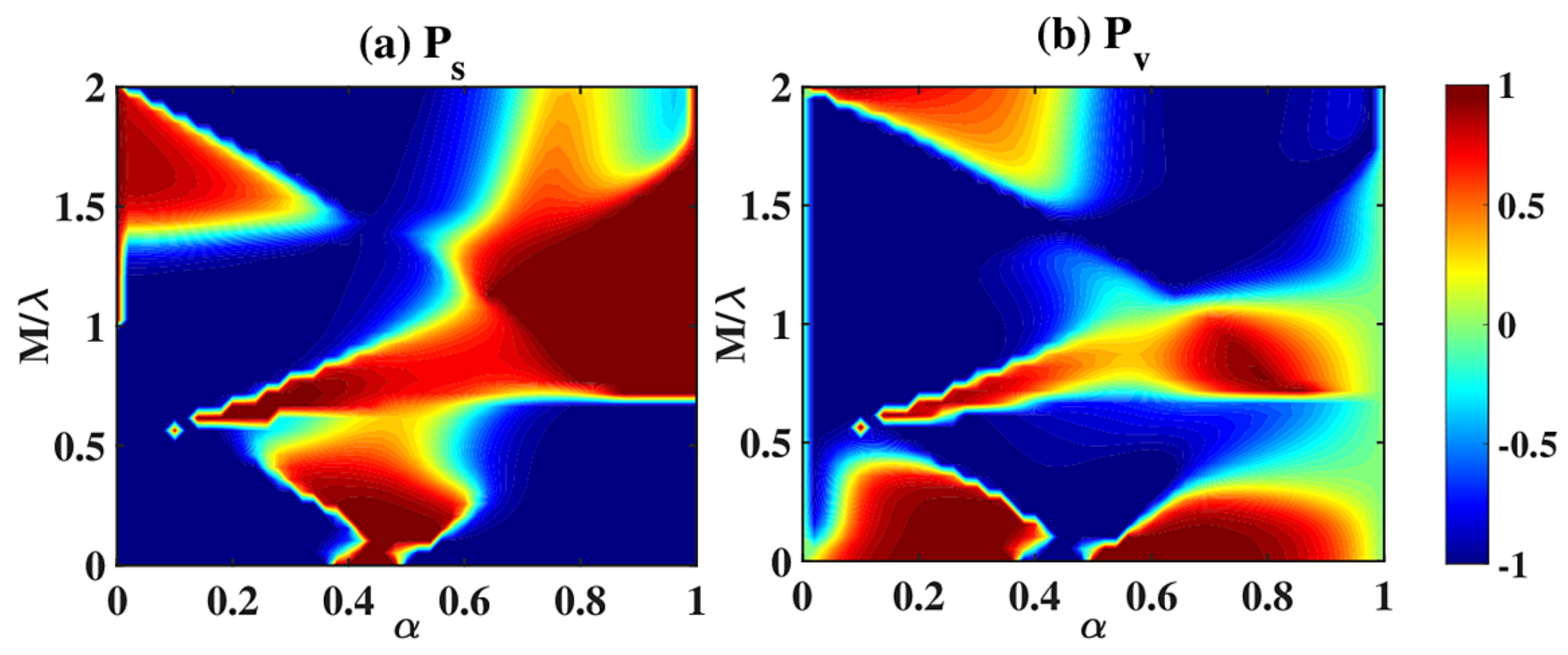}
\caption{(Color online) Contour plot of
(a) Spin polarization and (b) Valley polarization. We fix the chemical potential to $\mu=-50$ meV.}
\label{fig:spin__val_pol}
\end{figure}

In Fig.~\ref{fig:spin__val_pol}, we present the spin and valley polarizations over the entire range of $\alpha$, with the chemical potential fixed at $\mu=-50$ meV. 
The spin polarization $P_s$, shown in Fig.~~\ref{fig:spin__val_pol}(a), remains finite over a broad region of parameter space. As shown in the figure, $P_s$ exhibits several sign reversals and extended regions of nearly complete polarization $(|P_s|\approx1)$. At the two limiting points $\alpha=0$ and $\alpha=1$, the spin polarization exhibits sharp interfaces that arise from the vanishing of the corresponding effective mass terms. For graphene $(\alpha=0)$, the effective mass term is given by $\Delta_{\rm g}$, while for the dice lattice $(\alpha=1)$, it is given by $\Delta_{\rm d}$.
The conditions $\Delta_{\rm g}=0$ and 
$\Delta_{\rm d}=0$ therefore define the interfaces that separate two oppositely spin-polarized regimes. Physically, these conditions correspond to the sign reversal of the effective mass term, which changes the sign of the Berry curvature and consequently reverses the dominant spin contribution to the anomalous Nernst conductivity. For graphene, the transition between two oppositely spin-polarized regimes occurs near $M/\lambda=1$, whereas for the dice lattice the corresponding transition shifts to $M/\lambda=1/\sqrt{2}$.
In contrast, the valley polarization $P_v$, shown in Fig.~~\ref{fig:spin__val_pol}(b), displays qualitatively different behavior. Most notably, $P_v$ vanishes at the two limiting values $\alpha=0$ and $\alpha=1$, despite the breaking of time-reversal symmetry. The absence of valley polarization at these limits is due to the restoration of inversion symmetry. A finite valley polarization emerges only in the intermediate regime $0<\alpha<1$, where inversion symmetry is intrinsically broken. In this regime, the valley-resolved Berry curvatures and corresponding anomalous Nernst conductivities become inequivalent, leading to a finite valley polarization. The contour plot reveals extended regions with large positive and negative valley polarization, including nearly fully valley-polarized states.

We briefly address here the experimental relevance of the model parameters used in this study. The spin-orbit coupling strength used in our calculations is  $\lambda=100$ meV, which lies within the achievable range in engineered Dirac materials and proximity-engineered heterostructures. Several studies have demonstrated that spin-orbit coupling in Dirac materials can be substantially enhanced through the deposition of heavy adatoms, magnetic proximity effects, and hybridization with heavy-metal substrates~\cite{Weeks2011, Wang2015, Hallal2017, Marchenko2012}.
 Similarly, exchange fields of the order of several tens of meV can be generated through magnetic proximity effects or magnetic doping~\cite{Chang2013,Wei2013}. Consequently, the choice of the staggered magnetization value $M = 50$ meV is experimentally realistic.
 The temperature considered in our calculations, $T=20$ K, corresponds to the low-temperature regime routinely employed in spin-caloritronic and anomalous Nernst experiments ~\cite{Liu2020, Nayak2016}.

\section{Summary}\label{conclusion}

We have presented a comprehensive study of the spin- and valley-resolved anomalous Hall and Nernst effects in the spin-orbit coupled $\alpha$-$T_{3}$ system. By incorporating the Kane–Mele type SOI and staggered sublattice magnetization into the model, we have shown that the Berry curvature and entropy-weighted transport coefficients are highly tunable. 
In the TRS-preserving regime, the system exhibits a pure valley Hall effect accompanied by a robust spin Hall response. The corresponding Nernst conductivities display characteristic peak–dip structures governed by the entropy-weighted Berry curvature distribution near the band edges. For intermediate $\alpha$ values, these curvatures become sharply localized, resulting in enhanced Hall and Nernst signals.
When TRS is broken by the staggered magnetization, the transport behavior changes significantly. Valley degeneracy is lifted, producing strong asymmetries between the $K$ and $K^{\prime}$ valleys. The Berry curvature reorganizes into highly valley-polarized hotspots, generating large anomalous Hall and Nernst effects with tunable sign and magnitude. The strongest responses emerge at intermediate $\alpha$ values, where the interplay of flat band physics and staggered magnetization-induced band modification is most pronounced. We have also shown that the Nernst response of the $\alpha$-$T_3$ system can generate strong spin and valley polarizations, with a high degree of tunability controlled by the parameters $\mu$, $M$, $\lambda$, and  $\alpha$. Both the spin and valley polarizations exhibit rich structures, with extended regions of complete polarization. Overall, our results establish that Berry-curvature driven anomalous thermoelectric effects in the $\alpha$-$T_3$ system provide an efficient route to generate and control spin- and valley-polarized currents. 

The $\alpha$-$T_3$ lattice is primarily a synthetic, engineered platform that can be realized in systems such as cold-atom optical lattices and heterostructures. Our results demonstrate that the $\alpha$-$T_3$ lattice provides a theoretically tunable framework for exploring spin–valley caloritronic phenomena. However, these findings are intended to establish fundamental theoretical insights, and we do not suggest any immediate roadmap for practical device applications.

\appendix
\begin{widetext}
\section{Derivation of the Energy Eigenvalue Equation}\label{Appen_A}

The total Hamiltonian can be written as 
\begin{eqnarray}\label{Ham_tot}
H(\bm k)=H_\sigma^\eta(\bm k)+H_M=\begin{pmatrix}
\varepsilon_1 & {\rm g}_k\cos\phi & 0 \\
{\rm g}_k^*\cos\phi & \varepsilon_2 & {\rm g}_k\sin\phi \\
0 & {\rm g}_k^*\sin\phi & \varepsilon_3
\end{pmatrix},
\end{eqnarray}
where ${\rm g}_k=\hbar v_F(\eta k_x+ik_y)$ and the diagonal matrix elements are defined as
\begin{eqnarray}
\varepsilon_1
=\lambda\eta\sigma(-\cos\phi + m\eta),~~~
\varepsilon_2
=\lambda\eta\sigma(\cos\phi-\sin\phi),~~
\mbox{and}~~~
\varepsilon_3
=\lambda\eta\sigma(\sin\phi - m\eta),
\end{eqnarray}
where $m=M/\lambda$.

The energy eigenvalues \(\varepsilon\) are obtained from the secular equation
\begin{eqnarray}
\det[H(\bm k)-\varepsilon \mathbb{I}]=0,
\end{eqnarray}
where $\mathbb{I}$ is the $3\times3$ identity matrix.

Expanding the determinant leads to the cubic equation
\begin{eqnarray}\label{eq:cubic_general}
\varepsilon^3
-
t\,\varepsilon^2
+
p\,\varepsilon
+
q
=0,
\end{eqnarray}
where
\begin{eqnarray}
t=\varepsilon_1+\varepsilon_2+\varepsilon_3,
\end{eqnarray}
\begin{eqnarray}
p=
\varepsilon_1\varepsilon_3
+
\varepsilon_1\varepsilon_2
+
\varepsilon_2\varepsilon_3
-
|f_k|^2,
\end{eqnarray}
and
\begin{eqnarray}
q=
-\varepsilon_1\varepsilon_2\varepsilon_3
+
|f_k|^2
\left(
\varepsilon_1\sin^2\phi
+
\varepsilon_3\cos^2\phi
\right).
\end{eqnarray}

Substituting the explicit forms of
$\varepsilon_1$, $\varepsilon_2$, and $\varepsilon_3$, we obtain
\begin{equation}
t=0.
\end{equation}
The coefficient $p$ simplifies to
\begin{eqnarray}
p=
\frac{\lambda^2}{2}\sin(2\phi)
+
\sqrt{2}\lambda M\eta
\sin\!\left(\phi+\frac{\pi}{4}\right)
-
\left(
\lambda^2
+
\hbar^2 v_F^2 k^2
+
M^2
\right).
\end{eqnarray}
Similarly, the coefficient $q$ becomes
\begin{eqnarray}
q=
\frac{\lambda\eta\sigma}{\sqrt{2}}
\sin(2\phi)
\cos\!\left(\phi+\frac{\pi}{4}\right)
\left(
\lambda^2+\hbar^2 v_F^2 k^2
\right)
+
\sqrt{2}\lambda\eta\sigma M^2
\cos\!\left(\phi+\frac{\pi}{4}\right)
-
M\sigma\cos(2\phi)
\left(
\lambda^2+\hbar^2 v_F^2 k^2
\right).
\end{eqnarray}
Therefore, Eq.~\eqref{eq:cubic_general} is reduced to a depressed form $\varepsilon^3+p\varepsilon+q=0$. The standard Cardano-trigonometric solution gives the three real roots of the cubic depressed equation as:
\begin{eqnarray}
  \varepsilon^n_{\eta,\sigma}(\bm{k})
  = 2\sqrt{\frac{-p}{3}}\,
    \cos\!\left[\frac{1}{3} 
    \arccos\!\left(\frac{3q}{2p}\sqrt{\frac{-3}{p}}\right)
    -\frac{2\pi n}{3}\right],
    \quad n=0,1,2.
\end{eqnarray}

\section{Analytical derivation of the Berry curvature at \texorpdfstring{$k=0$}{k=0}}\label{Appen_B}

At $\bm k=0$, the Hamiltonian given in Eq.~\eqref{Ham_tot} becomes
\begin{eqnarray}\label{Ham_tot0}
H_\sigma^\eta(0)=\begin{pmatrix}
\varepsilon_1 & 0 & 0 \\
0 & \varepsilon_2 & 0 \\
0 & 0 & \varepsilon_3
\end{pmatrix},
\end{eqnarray}
The energy eigenvalues are simply $\varepsilon_1$, $\varepsilon_2$, and $\varepsilon_3$. The corresponding eigenspinors are respectively given by
\begin{eqnarray}
u_1=\begin{pmatrix}
    1\\
    0\\
    0
\end{pmatrix},~~~
u_2=\begin{pmatrix}
    0\\
    1\\
    0
\end{pmatrix},~~~\mbox{and}~~~
u_3=\begin{pmatrix}
    0\\
    0\\
    1
\end{pmatrix}.
\end{eqnarray}

The gauge-invariant expression of the Berry curvature is given by
\begin{eqnarray}\label{Berry_Cg}
{{\Omega}}^{n}_{\eta,\sigma}({\bm k})=-2\hbar^2\, {\rm Im}\sum\limits_{\substack{n^\prime\\
(n\neq n^\prime)}}
{\frac{{\langle u^n_{\eta,\sigma}\vert v_x\vert{u^{n^\prime}_{\eta,\sigma}}\rangle\langle u^{n^\prime}_{\eta,\sigma}\vert v_y \vert{u^n_{\eta,\sigma}}\rangle}}{\left[\varepsilon^{n}_{\eta,\sigma}(\bm k)-\varepsilon^{n^\prime}_{\eta,\sigma}(\bm k)\right]^2}},
\end{eqnarray}
where the velocity operators are defined as 
$v_i=\frac{1}{\hbar}\frac{\partial H(\bm k)}{\partial k_i}$ with $i=x,y$. They are calculated as
\begin{eqnarray}
v_x=\eta v_F\begin{pmatrix}
    0 & \cos\phi & 0\\
 \cos\phi & 0 & \sin\phi\\
    0 & \sin\phi & 0
\end{pmatrix}~~~\mbox{and}~~~v_y=v_F\begin{pmatrix}
    0 & i\cos\phi & 0\\
 -i\cos\phi & 0 & i\sin\phi\\
    0 & -i\sin\phi & 0
\end{pmatrix}.   
\end{eqnarray}
Here, $\vert u_{\eta,\sigma}^n\rangle \equiv u_j$ and $\varepsilon_{\eta,\sigma}^n\equiv \varepsilon_j$ with $j=1,2,3$. Using Eq.~\eqref{Berry_Cg}, it is straightforward to calculate the corresponding Berry curvature at $\bm k=0$ as
\begin{eqnarray}
 \Omega_1=2\eta\hbar^2 v_F^2\frac{\cos^2\phi}{\Delta_{12}^2},~~~
 \Omega_2=-2\eta\hbar^2 v_F^2\Bigg[\frac{\cos^2\phi}{\Delta_{12}^2}-\frac{\sin^2\phi}{\Delta_{23}^2}\Bigg],~~~\mbox{and}~~~
  \Omega_3=-2\eta\hbar^2 v_F^2\frac{\sin^2\phi}{\Delta_{23}^2},
\end{eqnarray}
where $\Delta_{ij}=\varepsilon_i-\varepsilon_j$ is band gap between $\varepsilon_i$ and $\varepsilon_j$.
It is therefore evident that the Berry curvature of a given band $\varepsilon_j$ scales inversely with the square of the energy gaps between that band and its adjacent bands. The relevant band gaps are
\begin{eqnarray}
\Delta_{12}=\lambda\eta\sigma(\sin\phi-2\cos\phi+m\eta)~~~~\mbox{and}~~~\Delta_{23}=\lambda\eta\sigma(\cos\phi-2\sin\phi+m\eta).\nonumber 
\end{eqnarray}
The square of $\Delta_{12}$ and $\Delta_{23}$ depends on $\alpha$ (or $\phi$) and $\eta$. Therefore, for a given 
$\lambda$, the interplay of 
$\alpha$ and staggered magnetization would significantly modify Berry curvature distribution in the valleys.

The energy eigenvalues $\varepsilon_1$, $\varepsilon_2$, and $\varepsilon_3$ can be identified as the CB, FB, and VB depending on the values of $\eta$, $\sigma$, $\alpha$, and $m$. For example, when $m=0$ \cite{Orbit_M}, they becomes $\varepsilon_1
=-\lambda\eta\sigma\cos\phi$, $\varepsilon_2
=\lambda\eta\sigma(\cos\phi-\sin\phi)$, and 
$\varepsilon_3=\lambda\eta\sigma\sin\phi$. For $(\eta,\sigma)=(K,\uparrow)$, 
$\varepsilon_2\rightarrow{\rm CB},~ \varepsilon_3\rightarrow{\rm FB},$ and $\varepsilon_1\rightarrow {\rm VB}$ when 
$\alpha<1/2$. On the other hand, $\varepsilon_3\rightarrow{\rm CB},~ \varepsilon_2\rightarrow{\rm FB},$ and $\varepsilon_1\rightarrow {\rm VB}$ when $\alpha>1/2$. Moreover, $\varepsilon_2$ and 
$\varepsilon_3$ becomes degenerate at 
$\alpha=1/2$ and corresponding Berry curvatures diverge. As $\alpha$ is tuned across this critical value, a band inversion between $\varepsilon_2$ and $\varepsilon_3$ takes place, indicating the onset of a TPT. For $m<1/\sqrt{2}$, similar band inversion occurs upon crossing the $m$-dependent  critical values $\alpha_\pm$ defined in Eq.~\eqref{alph_bn}.

\end{widetext}

\end{document}